\documentclass[aps,prd,nofootinbib,reprint,superscriptaddress,longbibliography]{revtex4-1}
\usepackage{etoolbox}
\usepackage{dcolumn}% Align table columns on decimal point
\usepackage{bm}% bold math
\usepackage{graphics}
\usepackage{xcolor}
\usepackage{dirtytalk}
\usepackage{graphicx} %Include figure filesusepackage{graphicx} %Include figure files
\usepackage{blindtext}
\usepackage{amsmath}
\usepackage{amsfonts}
\usepackage{amssymb}
\usepackage{mathtools,halloweenmath}
\usepackage[%
colorlinks=true,
urlcolor=blue,
linkcolor=red,
citecolor=blue
]{hyperref}
\usepackage{physics}
\NewDocumentCommand{\tens}{t_}
{%
	\IfBooleanTF{#1}
	{\tensop}
	{\otimes}%
}
\NewDocumentCommand{\tensop}{m}
{%
	\mathbin{\mathop{\otimes}\displaylimits_{#1}}%
}
\usepackage{natbib}
\begin{document}
\title{\bf Quantum chaos in the presence of non-conformality}
\vskip 1cm
\author{Ashis Saha}
\email{ashis.saha@bose.res.in}
\affiliation{Department of Astrophysics and High Energy Physics,\linebreak
	S.N.~Bose National Centre for Basic Sciences,\linebreak
	 JD Block, Sector-III, Salt Lake, Kolkata 700106, India}
\author{Sunandan Gangopadhyay}
\email{sunandan.gangopadhyay@bose.res.in}
\affiliation{Department of Astrophysics and High Energy Physics,\linebreak
	S.N.~Bose National Centre for Basic Sciences,\linebreak
	 JD Block, Sector-III, Salt Lake, Kolkata 700106, India}	

	\begin{abstract}
		\noindent The behaviour of a chaotic system and its effect on existing quantum correlation has been holographically studied in presence of non-conformality. Keeping in mind the gauge/gravity duality framework, the non-conformality in the dual field theory has been introduced by considering a Liouville type dilaton potential for the gravitational theory. The resulting black brane solution is associated with a parameter $\eta$ which represents the deviation from conformality. The parameters of chaos, namely, the Lyapunov exponent and butterfly velocity are computed by following the well-known shock wave analysis. The obtained results reveal that presence of non-conformality leads to suppression of the chaotic nature of a system. Further, for a particular value of the non-conformal parameter $\eta$, the system achieves Lyapunov stability resulting from the vanishing of both the Lyapunov exponent and as well as butterfly velocity. Interestingly, this particular value of $\eta$ matches with the previously given upper bound of $\eta$ known as Gubser bound in the literature. The effects of chaos and non-conformality on the existing correlation of a thermofield doublet state have been quantified by holographically computing the thermo mutual information in both the presence and absence of the shock wave. Furthermore, the entanglement velocity is also computed and the effect of non-conformality on it has been observed. Finally, the obtained results for the Lyapunov exponent and the butterfly velocity have also been computed from the pole-skipping analysis. The results from the two approaches agree with each other.
	\end{abstract}
\maketitle
\noindent Study of various properties of a chaotic system has always been a matter of great interest both theoretically and experimentally. For a classical system, the characterization of chaos is done with help of a parameter known as the Lyapunov exponent $\lambda_L$ which can be defined in the following way
\begin{eqnarray}\label{eq1}
\lambda_L\sim\frac{1}{t}\ln\left(\frac{\delta \xi(t)}{\delta\xi_0}\right)
\end{eqnarray}
where $\delta\xi(t)$ denotes the change in the phase-space trajectory of a classical dynamical system due to a change in the initial condition $\delta\xi_0$. The above relation also advocates for the fact that chaos is highly sensitive to the initial conditions. On the other hand, the study of chaos in case of a quantum many body system is quite non-trivial \cite{Stockmann_1999}. Traditionally, one characterizes chaos for a non-classical system by comparing its energy spectrum to the spectrum of random matrices \cite{Ullmo2014}. Apart from this approach, another way to probe quantum chaos is to compute the \textit{double commutator} of two generic local hermitian operators $\hat{V}(0)$ and $\hat{W}(x,t)$. This can be written down in the following way \cite{Larkin1969QuasiclassicalMI}
\begin{eqnarray}\label{eq2}
\mathcal{C}(x,t)&=&-\langle\left[\hat{W}(x,t),\hat{V}(0)\right]^2\rangle_{\beta}\nonumber\\
&=&-\langle \hat{W}(x,t)\hat{V}(0)\hat{W}(x,t)\hat{V}(0)\rangle_{\beta}~.
\end{eqnarray}
The above quantity $\mathcal{C}(x,t)$ measures how much effect the  perturbation $\hat{V}(0)$ at an earlier time creates on the later measurement of $\hat{W}(x,t)$. In other words, one intends to study at what rate the information gets transferred between two space-like separated points. This property leads to the phenomenon \textit{velocity of the butterfly effect} or commonly known as the \textit{butterfly velocity} \cite{Shenker:2013pqa,Roberts:2016wdl}. The butterfly velocity is a state dependent quantity and can be understood as the low energy analogue of the \textit{Lieb-Robinson velocity} \cite{Lieb:1972wy}. This implies that it acts as a bound for the rate of transfer of information for a quantum mechanical system at low energy scale. For large-$N$ gauge theories, the four-point correlator has the following form which stands to be very crucial for the study of quantum chaos \cite{Shenker:2013pqa,Shenker:2014cwa,Roberts:2014isa}
\begin{eqnarray}\label{eq3}
\mathcal{C}(x,t)\sim\exp[\lambda_L\left(t-t_*-\frac{|x|}{v_B}\right)]
\end{eqnarray}
where $\lambda_L$ is the Lyapunov exponent, $v_B$ is the butterfly velocity. It is worth noting that in the above relation, $\lambda_L$ is sometimes said to be the quantum mechanical analogue of the classical Lyapunov exponent \cite{Stanford:2015owe} which characterizes the growth of quantum chaos. The Lyapunov exponent satisfies the well-known MSS (Maldacena-Shenker-Stanford) bound $\lambda_L\leq \frac{2\pi}{\beta}$, where, $\beta$ is the inverse of Hawking temperature \cite{Maldacena:2015waa}. The bound is only saturated for maximally chaotic systems. Furthermore, the relation (given in eq.\eqref{eq3}) is true as long as $t_{dis}\ll t <t_*$, where, $t_{dis}$ is the \textit{dissipation time} which controls the late time behaviour of $\mathcal{C}(x,t)$ and $t_*$ is the \textit{scrambling time} at which $\mathcal{C}(x,t)$ becomes $\sim\mathcal{O}(1)$ \cite{Jahnke:2018off}. The scrambling time basically denotes the time-scale at which the given perturbation gets distributed among all the degrees of freedom of the chaotic quantum mechanical system. From eq.\eqref{eq3}, one can observe that the space-like separation between the operators further delays the scrambling of information in the system. On the other hand, the butterfly velocity characterizes the growth of the given perturbation $\hat{V}(0)$. This motivates one to define a \textit{butterfly effect light-cone} $t-t_*=\frac{|x|}{v_B}$ for the double commutator given in eq.\eqref{eq2}. Inside the cone ($t-t_*>\frac{|x|}{v_B}$), $\mathcal{C}(x,t)\sim\mathcal{O}(1)$, and outside the cone ($t-t_*<\frac{|x|}{v_B}$), $\mathcal{C}(x,t)\approx0$.\\

\noindent The initial motivation to study chaos in a holographic set up lies in the understanding that black holes are intrinsically thermal systems which are characterized by the Hawking temperature and it is a well-known fact that thermal systems are the primary playgrounds for chaos. Further, it has been noted that for black holes, the MSS bound is always saturated \cite{Maldacena:2015waa,Shenker:2013pqa} which depicts the fact that the black holes are always maximally chaotic or in other words, they are the fastest scramblers \cite{Sekino:2008he,Lashkari:2011yi}. Keeping this observation in mind along with the gauge/gravity duality \cite{Maldacena:1997re,Gubser:1998bc,Witten:1998qj}, one might propose the following. The holographic description\footnote{Existence of a one spatial dimension higher gravity solution is said to be the holographic description  of a control field theory in one spatial dimension less.} of a maximally chaotic finite temperature large-$N$ gauge theory is possible as long as the Lyapunov exponent saturates the MSS bound. Another interesting fact that we would like to mention is the following. Inclusion of higher-curvature corrections on the gravity side does not modify the MSS bound, that is, the bound still remains $\lambda_L=\frac{2\pi}{\beta}$ \cite{Roberts:2014isa}. However, the presence of higher-curvature corrections do change the butterfly velocity \cite{Roberts:2014isa,Grozdanov:2018kkt,Natsuume:2019vcv}. The conventional approach to holographically study quantum chaos relies on the dual description of the thermofield doublet (TFD) state \cite{Israel:1976ur,Maldacena:2001kr}. This consists of two completely disjoint quantum mechanical systems $\mathrm{QFT}_{L}$ and $\mathrm{QFT}_{R}$ along with their energy eigenstates denoted as $\ket{\mathrm{n}}_L$ and $\ket{\mathrm{n}}_R$. In this set up, the TFD state can be defined as
\begin{eqnarray}\label{eq4}
\ket{\mathrm{TFD}}=\frac{1}{\sqrt{Z}}\sum_{n}e^{-\frac{\beta E_n}{2}} \ket{\mathrm{n}}_L \tens \ket{\mathrm{n}}_R~.
\end{eqnarray}
\begin{figure}[!h]
	\centering
	\includegraphics[width=0.4\textwidth]{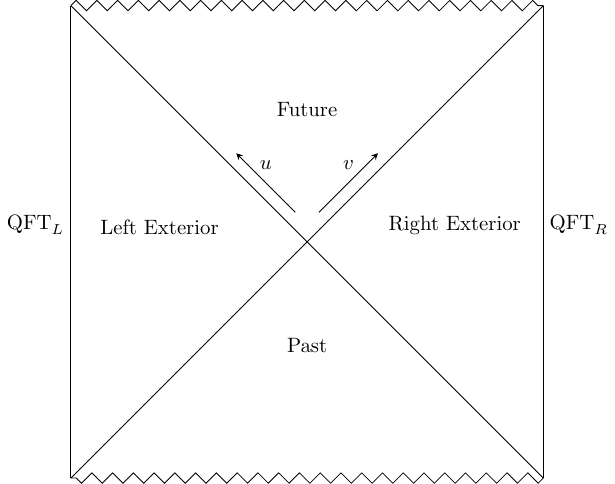}
	\caption{Dual geometry of the thermofield doublet state.}
	\label{TFD}
\end{figure}
The holographic description of the TFD state is a two-sided eternal black hole geometry in $AdS$ spacetime where the two black hole geometries are connected with each other by a non-traversable wormhole. The quantum theories are living on the two asymptotic boundaries (left and right) of the black hole spacetimes (a simple visualization of this set up has been given in Fig.\eqref{TFD}). As we mentioned earlier, the associated geometries of the asymptotic boundaries are connected with each other via a non-traversable wormhole and this ensures that the quantum theories (living on these boundaries) do not interact with each other. Entanglement is the sole reason due to which they are aware of each other. The TFD state implies that if we consider subsystems, namely, $A$ and $B$ which belongs to the systems $\mathrm{QFT}_{L}$ and $\mathrm{QFT}_{R}$ respectively then there exists a non-vanishing local correlation between $A$ and $B$. One can quantify this correlation with the help of the holographic thermo mutual information (HTMI) $I(A:B)$ which has the following definition \cite{Morrison:2012iz}
\begin{eqnarray}\label{eq5}
I(A:B) = S_{\mathrm{vn}}(A)+S_{\mathrm{vn}}(B)-S_{\mathrm{vn}}(A\cup B)
\end{eqnarray}
where $S_{\mathrm{vn}}(.)$ denotes the von Neumann entropy corresponding to the relevant subsystem. The holographic thermo mutual information is the generalization of the holographic mutual information \cite{Wolf:2007tdq,Hayden:2011ag,Fischler:2012uv} which was first introduced in \cite{Morrison:2012iz}. The computation of HTMI requires a wormhole which connects the asymptotic boundaries of a two-sided eternal black hole geometry in AdS. Similar to HMI, the HTMI is also UV finite and positive definite quantity. In the absence of chaos, HTMI is non-zero and positive which depicts the signature of entanglement between the subsystems $A$ and $B$ \cite{Wolf:2007tdq,Morrison:2012iz,Fischler:2012uv}. In order to observe the butterfly effect one needs to disrupt the existing local correlation between the subsystems of two copies of decoupled quantum systems. In other words, one has to disrupt the structure of TFD given in eq.\eqref{eq4}. In a holographic set up, this can be done by adding a small perturbation to the system in the asymptotic past. From the bulk perspective, the energy of the small perturbation (which has been added to the system at an early enough time) gets blue-shifted and falls to the black hole which results in the shock wave modification of the existing holographic geometry \cite{Dray:1984ha,Sfetsos:1994xa}. The mentioned process of perturbing the TFD structure disrupts the previously existing local correlation between the subsystems and then one can follow the approach shown in \cite{Shenker:2013pqa,Leichenauer:2014nxa} in order to obtain the Lyapunov exponent and butterfly velocity. Some of the recent works in this direction can be found in \cite{Sircar:2016old,Jahnke:2017iwi,Huang:2018snb,Avila:2018sqf,Fischler:2018kwt,Ageev:2021xgy,Mahish:2022xjz,Chakrabortty:2022kvq}. It is to be noted that the mentioned perturbation disrupts the correlation between the two decoupled theory living at the two boundaries of the two-sided eternal black hole geometry. In other words, it will only affect the term $S_{\mathrm{vn}}(A\cup B)$, not the individual entanglement pattern $S_{\mathrm{vn}}(A)$ or $S_{\mathrm{vn}}(B)$. Furthermore, the disruption of this two-sided correlation is controlled by the \textit{entanglement velocity} $v_{en}$ which basically probes the linear growth of $S_{\mathrm{vn}}(A\cup B)$ for ($t_e\geq t_*$) in the following way \cite{Hartman:2013qma,Hartman:2015apr,Mezei:2016zxg}
\begin{eqnarray}
\frac{d S_{\mathrm{vn}}(A\cup B)}{dt_e}=v_{en}S_{th}\mathrm{Area}_{\Sigma}
\end{eqnarray}
where $t_e$ denotes the time at which the perturbation is added to the system and $\mathrm{Area}_{\Sigma}=\partial(A\cup B)$. Further, $S_{th}$ is the density of the thermal entropy (Bekenstein-Hawking entropy of the black hole). This behaviour has been noted in both purely field theoretic set up \cite{Calabrese:2005in,Cotler:2016acd} and holographic set up and it can be explained with the help of the \textit{entanglement tsunami} phenomena \cite{Liu:2013iza,Liu:2013qca}. In \cite{Mezei:2016zxg}, it has been argued that the entanglement velocity is always bounded from above by the butterfly velocity for any holographic theory obeying the null energy condition (NEC), that is,
\begin{eqnarray}
v_{en}\leq v_B~.
\end{eqnarray}
Further, in \cite{Mezei:2016wfz}, it was proven that the above relation holds for any unitary quantum system. It is to be noted that both the butterfly velocity and entanglement velocity are always less than the speed of light \cite{Qi:2017ttv,Mezei:2016wfz} due to causality.\\
On other hand, recently, it was shown that in case of a quantum many body system, properties of a chaotic system can be characterized with the help of the energy density retarded Green's function \cite{Blake:2018leo,Grozdanov:2018kkt,Blake:2019otz}. In the context of AdS/CFT duality, the mentioned Green's function can be written down in the following form \cite{Natusuume,nastase}
\begin{eqnarray}\label{eq6}
G_{T^{00} T^{00}}^R=\frac{b(\omega,k)}{a(\omega,k)}~.
\end{eqnarray}
Now, the ``pole-skipping'' phenomena \cite{Grozdanov:2017ajz,Grozdanov:2019uhi} states that at some special points of complex $(\omega,k)$-plane, $b(\omega_*,k_*)=a(\omega_*,k_*)=0$ where $(\omega_*,k_*)$ denotes the mentioned special point. This implies at these special points, one has line of zeroes intersecting with the line of poles for the retarded Green's function. Further, this implies that at these points the retarded Green's function is non-unique or ill-defined. We follow the literature and denote these points as the pole-skipping points. For theories with holographic duals, the pole-skipping points can be obtained from the bulk field equations \cite{Blake:2019otz,Blake:2018leo,Natsuume:2019sfp,Natsuume:2019xcy}. In the holographic set up, the non-uniqueness of the retarded Green's function corresponding to the boundary theory is mapped to the non-uniqueness of the ingoing mode of the bulk field at the event horizon. It has been observed that for a static black hole, the pole-skipping frequency (leading order) is given by \cite{Ceplak:2019ymw,Ceplak:2021efc,Wang:2022mcq,Ning:2023ggs}
\begin{eqnarray}\label{eq7}
\omega_*=2\pi i T(s-1)
\end{eqnarray}
where $s$ represents the spin of the field operator. The above relation implies that depending upon the spin of the field, the position of the pole-skipping frequency varies in the complex-$\omega$ plane. Further, it has been observed that for strongly coupled theories with holographic duals, the leading order pole-skipping point located in the upper-half of the complex-$\omega$ plane is related to the Lyapunov exponent and the Butterfly velocity in the following way \cite{Grozdanov:2018kkt,Blake:2017ris,Blake:2018leo}
\begin{eqnarray}\label{eq8}
\omega_*=i\lambda_L;~k_*=\frac{i\lambda_L}{v_B}\equiv\frac{\omega_*}{v_B}~.
\end{eqnarray}
However, the above observation is only true for strongly coupled theories with holographic dual which are maximally chaotic. For non-maximally chaotic (strongly coupled) systems this is not true as only for the maximally chaotic systems, the stress tensor dominates chaos and pole-skipping only recovers the contributions of the stress tensor to the Lyapunov exponent and Butterfly velocity \cite{Choi:2020tdj}. From the subsequent discussion, we shall find that the results in this paper are compatible with the conclusions drawn in \cite{Choi:2020tdj}. The above relation of the Lyapunov exponent together with the relation given in eq.\eqref{eq7} depicts the fact that only for $s=2$ field (metric fluctuation) one gets the pole-skipping points in the upper-half of the complex-$\omega$ plane, which are related to the parameters of chaos. Furthermore, one can also observe that for $s=2$, one gets a maximally chaotic system as in this case $\lambda_L=2\pi T$ (saturated MSS bound). On the other hand, for other fields ($s=0,\frac{1}{2},1$, etc.), the pole-skipping point is not related to the parameters of chaos although the retarded Green's function is still non-unique at the pole-skipping points \cite{Ceplak:2021efc}. Some interesting works in this direction can be found in \cite{Ceplak:2019ymw,Natsuume:2019vcv,Abbasi:2019rhy,Ahn:2020bks,Ahn:2020baf,Kim:2020url,Natsuume:2020snz,Abbasi:2020xli,Choi:2020tdj,Ramirez:2020qer,Abbasi:2020ykq,Sil:2020jhr,Yuan:2020fvv,Blake:2021hjj,Ceplak:2021efc,Yuan:2021ets,Wang:2022mcq,Yuan:2023tft,Yadav:2023hyg, Ahn:2019rnq,Jeong:2021zhz,Jeong:2022luo,Jeong:2023rck,Jeong:2023ynk}.\\
In this work we consider the black brane solution of the Einstein-dilaton theory with a Liouville type profile for the dilaton potential \cite{Kulkarni:2012re,Kulkarni:2012in,Park:2012lzs,Park:2015afa} as the gravitational theory. The motivation to consider such theory lies in the fact that in the asymptotic limit we get a warped geometry instead of a $AdS$ geometry. This in turn means that in the context of gauge/gravity duality, the boundary field theory will be relativistic but nonconformal in nature \cite{Charmousis:2009xr}. This type of geometry belongs to the class of geometries (such as Lifshitz geometry, hyperscaling violating geometry, anisotropic geometry, etc.) which help us to generalize the gauge/gravity duality. The geometry under consideration is characterized by the parameter $\eta$. In the limit $\eta\rightarrow0$, we obtain the Schwarzschild type black brane solution. This nonconformal parameter also satisfies a certain bound known as the Gubser bound \cite{Gubser:2000nd,Gouteraux:2011ce}.\\
The plan of this paper is as follows. In section \eqref{section1}, we briefly discuss about the holographic gravitational dual of the non-conformal theory. We then compute the corresponding shock wave geometry and compute the Lyapunov exponent and butterfly velocity in section \eqref{section2}. In section \eqref{section3}, we quantify the effect of shock wave on the holographic thermo mutual information and also compute the entanglement velocity. We once again compute the Lyapunov exponent and butterfly velocity from the lowest order pole-skipping points. This we provide in section \eqref{section4}. In this section, we also compute the higher order pole-skipping points by considering scalar field fluctuations. We summarize our findings in section \eqref{section5}. 
\section{Einstein-dilaton theory with Liouville potential}\label{section1}
\noindent The Einstein-Hilbert action corresponding to the ($d+1$)-dimensional Einstein-dilaton theory with Liouville type potential reads \cite{Kulkarni:2012re,Kulkarni:2012in,Park:2012lzs,Park:2015afa}
\begin{eqnarray}\label{bulkaction}
S_{\mathrm{EH}}=\frac{1}{16\pi G_N^{d+1}}\int d^{d+1}x \sqrt{-g}\left[R-2(\partial\phi)^2-V(\phi)\right]\nonumber\\
\end{eqnarray}
where $V(\phi)=2\Lambda e^{\eta\phi}$ is the Liouville type dilaton potential.  Here $\Lambda$ represents the cosmological constant $\Lambda<0$ and $\eta$ denotes the non-conformal parameter which captures the deviation of the system from conformality. Further, $\eta<\sqrt{\frac{8d}{(d-1)}}$ \cite{Gubser:2000nd,Gouteraux:2011ce}. The corresponding Einstein field equations and the equation of motion for the dilaton field reads
\begin{eqnarray}\label{EOM}
R_{\mu\nu}-\frac{1}{2}g_{\mu\nu}R+\frac{1}{2}g_{\mu\nu}V(\phi)&=&2\partial_{\mu}\phi\partial_{\nu}\phi-g_{\mu\nu}(\partial\phi)^2\nonumber\\
\frac{1}{\sqrt{-g}}\partial_{\mu}\left(\sqrt{-g}g^{\mu\nu}\partial_{\nu}\phi\right)&=&\frac{1}{4}\frac{\partial V(\phi)}{\partial \phi}~.
\end{eqnarray}
By solving the above two equations, one can obtain the following non-conformal black brane geometry\footnote{We have set $G_N=1$ and AdS radius $R=1$, for the sake of simplicity.} \cite{Kulkarni:2012re,Kulkarni:2012in,Park:2012lzs,Park:2015afa,Saha:2020fon}
\begin{eqnarray}\label{EdBB}
	ds^2&=&-r^{2p}f(r) dt^2+\frac{dr^2}{r^{2p}f(r)}+r^{2p}\sum_{i=1}^{d-1}dx_{i}^2\\
	f(r)&=&1-\left(\frac{r_+}{r}\right)^c\nonumber
\end{eqnarray}
where we have used
\begin{eqnarray}
p=\frac{8}{8+(d-1)\eta^2};~c=\frac{8d-(d-1)\eta^2}{8+(d-1)\eta^2}~.
\end{eqnarray}
As mentioned earlier, the asymptotic limit of the above given black brane is not $AdS$ and therefore the boundary field theory is not conformal. Further, it can be observed that in the limit $\eta\rightarrow0$, one obtains the $AdS_{d+1}$-Schwarzschild black brane solution. Recently, in \cite{Chew:2022enh,Chew:2023olq}, the authors have applied two different type of scalar potentials to numerically construct the solutions of hairy black holes and scalarons to study their properties systematically. These hairy black holes also bifurcated from the Schwarzschild black hole.\\
The Hawking temperature of the black brane geometry (given in eq.\eqref{EdBB}) reads
\begin{eqnarray}\label{HawkingT}
T_{H} = \frac{k}{2\pi}= \left(\frac{c}{4\pi}\right)r_+^{2p-1}
\end{eqnarray}
where $k$ is the surface gravity. It is to be observed that for $\eta=\sqrt{\frac{8d}{(d-1)}}$, the Hawking temperature of the black brane is zero, irrespective of the value of $r_+$. Furthermore, from the above relation, one can express the event horizon position $r_+$ in terms of the Hawking temperature. This reads
\begin{eqnarray}\label{EHinBeta}
r_+=\left(\frac{c}{4\pi T_H}\right)^{\frac{1}{1-2p}}=\left(\frac{\beta c}{4\pi}\right)^{\frac{1}{1-2p}}~.
\end{eqnarray}
\section{Shock wave analysis: Lyapunov exponent and butterfly velocity}\label{section2}
In this section, we now proceed to compute the Lyapunov exponent and butterfly velocity by carrying out the shock wave analysis. In order to obtain the shock wave geometry in this set up, we first write down the metric (given in eq.\eqref{EdBB}) in the Kruskal coordinates as it provides convenience in case of a two-sided geometry set up.\\
The first step is to introduce the Tortoise coordinate. For a general metric of the form
\begin{eqnarray}
	ds^2=-G_{tt}(r)dt^2+G_{rr}(r)dr^2+G_{ij}(r)dx^idx^j,
\end{eqnarray}
the Tortoise coordinate is defined as
\begin{eqnarray}
	dr_*=-\sqrt{\frac{G_{rr}(r)}{G_{tt}(r)}}dr~.
\end{eqnarray}
For the metric given in eq.\eqref{EdBB}, the above equation leads to
\begin{eqnarray}
dr_* =-\frac{dr}{r^{2p}f(r)}~.
\end{eqnarray}
With the above transformation in hand, we now move on to the following Kruskal coordinates
\begin{eqnarray}
u=e^{-k(t-r_*)},~v=-e^{k(t+r_*)}~.
\end{eqnarray}
By incorporating the Kruskal coordinate transformation, one obtains the following form
\begin{eqnarray}\label{EdBB2}
ds^2=2\Omega(u,v)dudv+g_{ij}(u,v)dx^idx^j
\end{eqnarray}
where 
\begin{eqnarray}
2\Omega(u,v)=\frac{\beta^2r^{2p}f(r)}{4\pi^2uv}~.
\end{eqnarray}
In Kruskal coordinates, the event horizon lies at $u=0$ or $v=0$. The exterior regions are located at $u>0,v<0$ (right exterior) or $u<0,v>0$ (left exterior). On the other hand, the singularity is located at $uv=1$ and the boundary is at $uv=-1$. In Fig.(\ref{TFD}), we have given the Penrose-Carter diagram depicting all the facts mentioned above. One can now assume the following general form of the stress-tensor \cite{Sfetsos:1994xa} corresponding to the unperturbed metric given in eq.\eqref{EdBB}
\begin{eqnarray}\label{matterS}
T^{matter}&=&T_{uu}du^2+T_{vv}dv^2+2T_{uv}dudv\nonumber\\
&&+T_{ij}dx^{i}dx^{j}
\end{eqnarray}
where the components $T_{uu}$, $T_{uv}$, $T_{vv}$ and $T_{ij}$ are functions of $u$ and $v$. This general form of the stress tensor will be used later as an ansatz in the backreacted shock wave geometry. Further, we assume that the metric given in eq.\eqref{EdBB2} is a solution of the following Einstein equation
\begin{eqnarray}\label{unperEnstn}
R_{ab}-\frac{1}{2}g_{ab}R=8\pi T_{ab}^{matter}
\end{eqnarray}
where the form of the components $T_{ab}^{matter}$ are given in eq.\eqref{matterS}. It is to be mentioned that the contribution coming form the cosmological constant is also taken care by the part $T_{ab}^{matter}$ \cite{Fischler:2018kwt}. Next, we consider that a tiny pulse of energy $E_0$ is added to left side of the geometry from the boundary at an earlier time $t$. Considering $t=0$ as the reference frame, the energy of the added perturbation (at earlier time $t$) gets blue-shifted and it follows an almost null trajectory towards the past horizon. This process introduces non-trivial modification to the original geometry \cite{Dray:1984ha,Sfetsos:1994xa}. Our motivation is to compute the change to the unperturbed geometry (given in eq.\eqref{EdBB2}) and the associated stress tensor (given in eq.\eqref{matterS}) due to the presence of the null pulse. In order to do this, we assume that the null pulse of energy is localized at $u=0$ and it propagates along the $v$-direction. In terms of the Penrose diagram, this consideration creates an extension along the $v$-direction. Mathematically, this extension can be incorporated by considering the following transformations
\begin{eqnarray}\label{perturbedKrs}
v\rightarrow v&+&\theta(u)\alpha(t,x^i)\nonumber\\
dv\rightarrow dv&+&\theta(u)\partial_i\alpha(t,x^i)dx^i
\end{eqnarray}
where the form of the function $\alpha(t,x^i)$ is to be determined from the Einstein field equations. 
\begin{figure}[!h]
	\centering
	\includegraphics[width=0.4\textwidth]{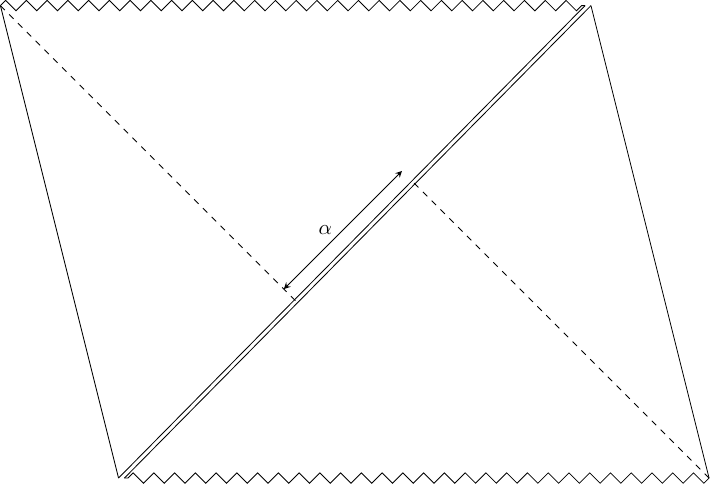}
	\caption{Penrose diagram of the shock wave geometry.}
	\label{SW}
\end{figure}
The theta function $\theta(u)$ ensures that the changes are constrained to the region $u>0$ (left exterior). As mentioned earlier, the said non-trivial modification to the original spacetime can be understood with the help of a Penrose diagram. This we have given in Fig.\eqref{SW}. We now incorporate the transformations given in eq.\eqref{perturbedKrs} to the unperturbed metric (given in eq.\eqref{EdBB2}) and obtain the following form for the backreacted geometry
\begin{eqnarray}\label{perturbed_metric}
ds^2&=&2\Omega(u,v+\theta(u)\alpha(t,x^i)) ~du ~(dv+\theta(u)\partial_i\alpha(t,x^i)dx^i)\nonumber\\
&&+g_{ij}(u,v+\theta(u)\alpha(t,x^i))dx^idx^j~.
\end{eqnarray}
For the sake of convenience, we now introduce the following set of new coordinates
\begin{eqnarray}
\bar{u}&=&u\nonumber\\
\bar{v}&=&v+\theta(u)\alpha(t,x^i)\nonumber\\
\bar{x}^i &=& x^{i}~.
\end{eqnarray}
From the above given coordinate transformations, one can easily show the following relation
\begin{eqnarray}\label{EqNew1}
dv+\theta(u)\partial_i\alpha(t,x^i)dx^i&=&d\bar{v}-\delta(u)\alpha(t,x^i)du\nonumber\\
&=&d\bar{v}-\delta(\bar{u})\alpha(t,\bar{x}^i)d\bar{u}~.
\end{eqnarray}
In terms of these new coordinates, the backreacted metric (given in eq.\eqref{perturbed_metric}) takes the following form
\begin{eqnarray}\label{perturbed2}
ds^2&=&2\Omega(\bar{u},\bar{v}) d\bar{u} [d\bar{v}-\delta(\bar{u})\alpha(t,\bar{x}^i)d\bar{u}]\nonumber\\
&&+g_{ij}(\bar{u},\bar{v})d\bar{x}^i d\bar{x}^j~.
\end{eqnarray}
In obtaining the above metric, we have used the relation given in eq.\eqref{EqNew1}. The above given metric is usually denoted as the general form of the shock wave metric. On the other hand, the general energy-momentum stress tensor corresponding to the matter part (given in eq.\eqref{matterS}) is also modified as
\begin{eqnarray}\label{matterS2}
T^{matter}=&&[\bar{T}_{uu}-2\delta(\bar{u})\alpha(t,\bar{x}^i)\bar{T}_{uv}+\delta^2(\bar{u})\alpha^2(t,\bar{x}^i)\bar{T}_{vv}]d\bar{u}^2\nonumber\\
&&+\bar{T}_{vv}d\bar{v}^2+2[\bar{T}_{uv}-\alpha(t,\bar{x}^i)\delta(\bar{u})\bar{T}_{vv}]d\bar{u}~d\bar{v}\nonumber\\
&&+\bar{T}_{ij}d\bar{x}^i~d\bar{x}^j
\end{eqnarray}
where $T_{ab}(u,v+\theta(u)\alpha(t,x^i))\equiv \bar{T}_{ab}$. Furthermore, the stress tensor associated to the shock wave is assumed to have the following form \cite{Shenker:2013pqa}
\begin{eqnarray}\label{matterSW}
T^{SW} = E_0 e^{\frac{2\pi}{\beta}t} \delta(\bar{u}) d\bar{u}^2\delta(\bar{x}^i)
\end{eqnarray}
where $E_0$ is the asymptotic energy of the null pulse, $e^{\frac{2\pi}{\beta}t}$ is the blue shift factor. Further $\delta(\bar{x}^i)$ ensures that the perturbation is localized at $x^i=0$. In order to find the profile of $\alpha(t,\bar{x}^i)$, we assume that the perturbed metric (given in eq.\eqref{perturbed2}) is a valid solution of the following Einstein equation
\begin{eqnarray}\label{pEq}
R_{ab}-\frac{1}{2}g_{ab}R=8\pi \left(T_{ab}^{matter}+T_{ab}^{SW}\right)
\end{eqnarray}
where the expressions of $T_{ab}^{matter}$ and $T_{ab}^{SW}$ are given in eq.\eqref{matterS2} and eq.\eqref{matterSW} respectively.
For the sake of simplicity, we now introduce a book-keeping parameter $\epsilon$ in $\alpha(t,\bar{x}^i)$ as, $\alpha(t,\bar{x}^i)\rightarrow\epsilon\alpha(t,\bar{x}^i)$ and $T_{ab}^{SW}$ as $T_{ab}^{SW}\rightarrow\epsilon T_{ab}^{SW}$, where $|\epsilon|\ll1$. This process helps us in recovering the unperturbed Einstein equation (given in eq.\eqref{unperEnstn}) in the limit $\epsilon\rightarrow0$.\\

\noindent Firstly, we solve the unperturbed Einstein field equation (given in eq.\eqref{unperEnstn}) in order to obtain the values of $T_{uu}$, $T_{vv}$ and $T_{uv}$. We then substitute these values in the $uu$-component of the perturbed Einstein field equation given in eq.\eqref{pEq} and keep terms upto $\sim~\mathcal{O}(\epsilon)$. We then observe that the shock wave parameter $\alpha(t,\bar{x}^i)$ satisfies the following equation (on the horizon $u=0$ or $r=r_+$)
\begin{eqnarray}\label{characteristicseq}
	\delta(\bar{u})g^{ij}\left[\Omega(\bar{u},\bar{v})\partial_i\partial_j-\frac{1}{2}g_{ij},_{\bar{u}\bar{v}}\right]\alpha(t,\bar{x}^i)=8\pi T^{SW}_{uu}~.~
\end{eqnarray}
The above equation can be obtained from the $uu$-component of the perturbed Einstein equation (given in eq.\eqref{pEq}). We would like to mention that in obtaining the above equation, the following conditions must hold \cite{Sfetsos:1994xa,Jahnke:2017iwi}
\begin{eqnarray}
\Omega(\bar{u},\bar{v}),_{\bar{v}}=g_{ij},_{\bar{v}}=T^{matter}_{vv}=0~\mathrm{at}~\bar{u}=0~.
\end{eqnarray}
We now proceed to express eq.\eqref{characteristicseq} in terms of the $t$ and $r$ coordinates in which the background spacetime (eq.\eqref{EdBB}) was initially expressed. Furthermore, one needs to keep in mind the fact that eq.\eqref{characteristicseq} is to be evaluated on the horizon $\bar{u}=0$ or $r=r_+$.This motivates us to consider the following near-horizon expansion of the black hole lapse function $\xi(r)\equiv r^{2p}f(r)$ 
\begin{eqnarray}\label{NH_expansion}
	\xi(r)&=& \xi(r_+)+\partial_r \xi(r)\bigg|_{r=r_+}(r-r_+)+....\nonumber\\
	&=&cr_+^{2p-1}(r-r_+)+...~.
\end{eqnarray}\\
In the above computation we have used the fact that $f(r)|_{r=r_+}=0$. By using the above near-horizon form, one can show that the Tortoise coordinate takes the following form
\begin{eqnarray}\label{Tortoise_NH}
	r_*\approx\left(\frac{1}{cr_+^{2p-1}}\right)\ln(\frac{r-r_+}{r_+})~.
\end{eqnarray}
Further, the expression of $uv$ is obtained to be in the near-horizon limit
\begin{eqnarray}
	uv= e^{\frac{4\pi}{\beta}r_*}\approx e^{\ln(\frac{r-r_+}{r_+})}~.
\end{eqnarray}
We now make use of the above near-horizon expressions to compute the expression of $\Omega(\bar{u},\bar{v})$ and $g_{ij},_{\bar{u}\bar{v}}$ on the horizon $\bar{u}=0$ which appear in eq.\eqref{characteristicseq}.\\
This reads
\begin{eqnarray}\label{NH_2}
	\Omega(\bar{u},\bar{v})\bigg|_{\bar{u}=0}&=&\frac{\beta^2r^{2p}f(r)}{8\pi^2uv}\bigg|_{\bar{u}=0}\nonumber\\
	&\approx&\frac{1}{2} \left(\frac{2}{cr^{2p-1}_+}\right)^2 cr^{2p-1}_+(r-r_+)e^{-\ln(\frac{r-r_+}{r_+})}+...\nonumber\\
	&\approx& \frac{2}{cr_+^{2p-2}}\equiv\Omega(r_+)~.
\end{eqnarray}
In the above computation, we have used the near-horizon expansions given in eq.(s)(\eqref{NH_expansion},\eqref{Tortoise_NH}).\\
We now proceed to compute $\frac{1}{2}\frac{dg_{ij}}{d(\bar{u}\bar{v})}\vert_{\bar{u}=0}$. This reads
\begin{eqnarray}
	\frac{dg_{ij}}{d(\bar{u}\bar{v})}\bigg|_{\bar{u}=0}&=&\left(\frac{dg_{ij}}{dr_*}\right)\left(\frac{dr_*}{d(\bar{u}\bar{v})}\right)\bigg|_{\bar{u}=0}\nonumber\\
	&=&r^{2p}f(r)\frac{\beta}{4\pi}e^{-\frac{4\pi r_*}{\beta}}\partial_r g_{ij}\bigg|_{r=r_+}
\end{eqnarray}
where $\beta=\frac{4\pi}{c r_+^{2p-1}}$ is the inverse Hawking temperature. Next, we use the near-horizon expansions given in eq.(s)(\eqref{NH_expansion},\eqref{Tortoise_NH}) in the above expression. This in turn gives us
\begin{eqnarray}\label{NH_1}
	\frac{dg_{ij}}{d(\bar{u}\bar{v})}\bigg|_{\bar{u}=0}&\approx&\left(cr_+^{2p-1}\right)\left(\frac{r-r_+}{cr_+^{2p-1}}\right)e^{-\ln(\frac{r-r_+}{r_+})}\partial_rg_{ij}\bigg|_{r=r_+}\nonumber\\
	&=&r_+\partial_rg_{ij}\bigg|_{r=r_+}~.
\end{eqnarray}
For the sake of simplicity, we have denoted $r_+\partial_r g_{ij}\vert_{r=r_+}$ as $r_+\partial_r g_{ij}(r_+)$.
\begin{widetext}
\noindent By using the above results in eq.\eqref{characteristicseq}, we obtain the following form	
\begin{eqnarray}\label{characteristicseqintr}
	g^{ij}(r_+)\left[\Omega(r_+)\partial_i\partial_j-\frac{r_+}{2}\partial_r g_{ij}(r_+)\right]\alpha(t,x^i)=8\pi E_0 e^{\frac{2\pi}{\beta}t}\delta(\bar{x}^i)~.\nonumber\\
\end{eqnarray}
We now choose the direction of propagation for the perturbation as $x^i=x_1$. This further simplifies the eq.\eqref{characteristicseqintr} to the following form
\begin{eqnarray}\label{eqtoSolve}
\left[\partial^2_{x_1}-M^2\right]\alpha(t,x_1)=E_0 e^{\frac{2\pi}{\beta}(t-t_*)}\delta(\bar{x}^i)
\end{eqnarray}
where
\begin{eqnarray}\label{Scram1}
M^2=\frac{(d-1)r_+}{2\Omega(r_+)}\partial_r g_{x_1x_1}(r_+),~t_*=\left(\frac{\beta}{2\pi}\right)\log(\frac{\Omega(r_+)}{8\pi g_{x_1x_1}(r_+)})~.
\end{eqnarray}
\end{widetext}
We now make use of the metric given in eq.\eqref{EdBB} along with eq.\eqref{NH_2} and the fact that $g_{x_1x_1}(r_+)=r^{2p}_+$ in order to obtain explicit expressions for $M^2$ and $t_*$. These read 
\begin{eqnarray}\label{explicit}
	M^2&=&\frac{c}{2}p(d-1)r_+^{2(2p-1)}\\
	t_*&=& \left(\frac{\beta}{2\pi}\right)\log(\frac{r_+^{p(d-1)}}{4\pi cr_+^{p(d+2)-2}})~.
\end{eqnarray}
In the above expression, $t_*$ represents the scrambling time which can be recast to the following standard form
\begin{eqnarray}
	t_*\approx\left(\frac{\beta}{2\pi}\right) \log(S_{\mathrm{BH}})
\end{eqnarray}
where $S_{\mathrm{BH}}$ is the Bekenstein-Hawking entropy density of the non-conformal black brane which has the form
\begin{eqnarray}
S_{\mathrm{BH}}=\frac{r_+^{p(d-1)}}{4}~.
\end{eqnarray}
We now proceed to solve eq.\eqref{eqtoSolve}. For $|x_1|\neq0$, the solution for $\alpha(t,x_1)$ reads
\begin{equation}\label{Cases}
	\alpha(t,x_1)=\begin{cases}
		c_0 e^{Mx_1};~\mathrm{for}~x<0\\
		c_1 e^{-Mx_1};~\mathrm{for}~x>0~.
	\end{cases}
\end{equation}
\noindent The above solution for both the regions must be continuous at $x=0$ which gives the condition $c_0=c_1$. On the other hand, by integrating eq.\eqref{eqtoSolve} one obtains
\begin{widetext}
\begin{eqnarray}
	\lim_{\epsilon\rightarrow0}\left[	\int_{x_1=0-\epsilon}^{x_1=0+\epsilon}\left(\partial^2_{x_1}-M^2\right)\alpha(t,x_1)dx_1\right]&=&E_0 e^{\frac{2\pi}{\beta}(t-t_*)}\lim_{\epsilon\rightarrow0}\left[\int_{x_1=0-\epsilon}^{x_1=0+\epsilon}\delta(x_1) dx_1\right]\nonumber\\
	\Rightarrow \lim_{\epsilon\rightarrow0}\left[\partial_{x_1}\alpha(t,x_1)\bigg|_{x_1=0+\epsilon}-\partial_{x_1}\alpha(t,x_1)\bigg|_{x_1=0-\epsilon}\right]&=&E_0e^{\frac{2\pi}{\beta}(t-t_*)}~.
\end{eqnarray}
\end{widetext}
We now make use of the solution given in eq.\eqref{Cases} in above equation which in turn gives
\begin{eqnarray}\label{SolutionEq_2}
	c_0+c_1=\frac{E_0}{M}e^{\frac{2\pi}{\beta}(t-t_*)}~.
\end{eqnarray}
By using the condition $c_0=c_1$ along with eq.\eqref{SolutionEq_2}, we obtain the solution for eq.\eqref{eqtoSolve} to be
\begin{eqnarray}
\alpha(t,x_1)&=&\frac{E_0}{2M} e^{\frac{2\pi}{\beta}(t-t_*)-M|x_1|}\nonumber\\
&\sim& \mathrm{constant} \times e^{\frac{2\pi}{\beta}(t-t_*)-M|x_1|}~.
\end{eqnarray}
\noindent A few comments are in order now. Keeping in mind eq.\eqref{eq2}, it is to be noted that the difference between states created by $\hat{W}(x,t)\hat{V}(0)$ and $\hat{V}(0)\hat{W}(x,t)$ is related to the null shift of the operator $\hat{V}(0)$ created by the shock wave profile $\alpha(x,t)$. As the commutator $\langle\left[\hat{W}(x,t),\hat{V}(0)\right]^2\rangle_{\beta}$ is determined by the real part of $\langle\hat{W}(x,t)\hat{V}(0)\hat{W}(x,t)\hat{V}(0)\rangle_{\beta}$ \cite{Roberts:2014isa,Shenker:2013yza}, at early times $\beta < t < t_*+\frac{x}{v_B}$, one has \cite{Shenker:2013yza,Roberts:2014isa,Roberts:2016wdl}
\begin{eqnarray}
	\mathcal{C}(t,x)\sim \alpha(t,x)~.
	\end{eqnarray}
We now compute the expressions for Lyapunov exponent and butterfly velocity by comparing the above with eq.\eqref{eq3}. This yields the following results
\begin{eqnarray}\label{result1}
\lambda_L=\frac{2\pi}{\beta};~v_B=\frac{1}{4}\sqrt{\frac{8d}{d-1}-\eta^2}
\end{eqnarray}
where the expression for $\beta$ is given in eq.\eqref{HawkingT}.
\begin{figure}[!h]
%	\begin{minipage}[t]{0.4\textwidth}
%		\centering\includegraphics[width=\textwidth]{Lyapunov.pdf}\\
%	\end{minipage}\hfill
	\begin{minipage}[t]{0.4\textwidth}
		\centering\includegraphics[width=\textwidth]{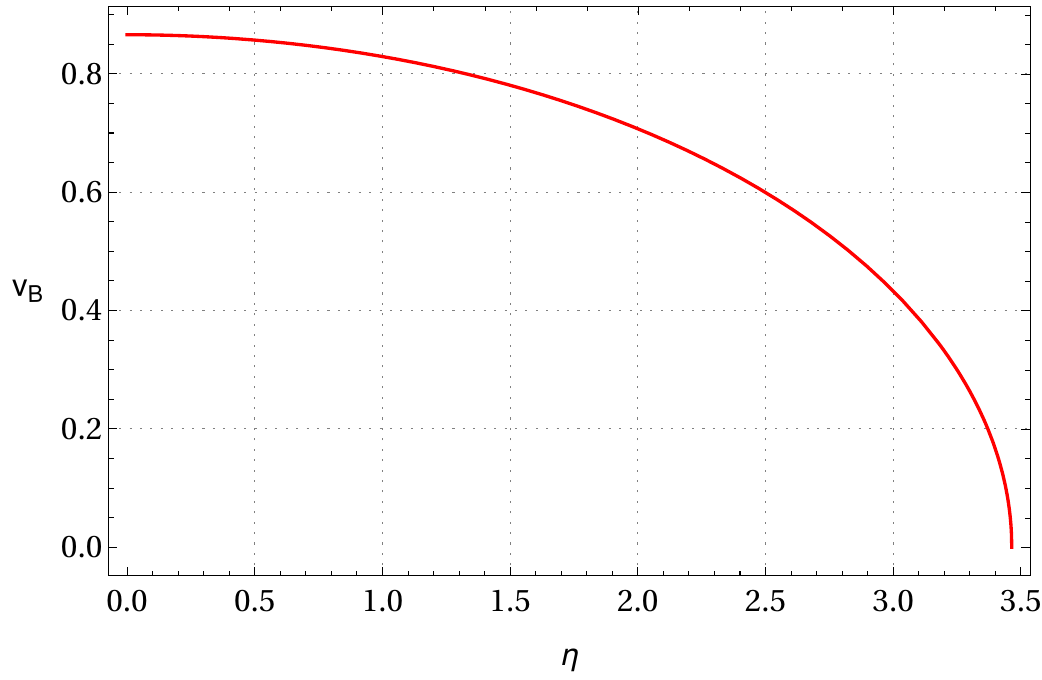}\\
	\end{minipage}
	\caption{Variation of the butterfly velocity ($v_B$) with respect to the non-conformal parameter $\eta$, where we set $d=3$.}
	\label{Butterfly}
\end{figure}
In the conformal limit $\eta\rightarrow0$, one obtains $\beta=\frac{dr_+}{4\pi}$ and $v_B=\sqrt{\frac{d}{2(d-1)}}$. This in turn means that in this particular limit one recovers the SAdS$_{d+1}$ results given in \cite{Shenker:2013pqa}. We would now like to make a few comments regarding our results. It can be observed that the form of the Lyapunov exponent remains unchanged, that is, $\lambda_L = 2\pi T_{H}$ and this form is universal for any maximally chaotic holographic theory. Further, it also depicts the fact that the Lyapunov exponent depends only on the Hawking temperature $T_H$ (which is also the temperature of the dual field theory). However, the Hawking temperature is subject to the holographic model under consideration which in our case corresponds to the non-conformal black brane solution. Here, the parameter $\eta$ in the gravitational solution characterizes the theory since it appears in the bulk action (eq.\eqref{bulkaction}). From the perspective of the dual theory, $\eta$ corresponds to turning on some deformation away from conformality, which implies choosing a particular boundary theory. Once $\eta$ is chosen, we would then fix $\beta$, which characterizes the thermal state of the dual field theory, to make plots of the various quantities that we shall compute in the subsequent sections. 	
%In general, by choosing a value for the position of the event horizon $r_+$ (which in turn fixes the value of the temperature of the dual field theory), one obtains the value of the Lyapunov exponent at a particular temperature. For the geometry in hand, that is, for the non-conformal black brane solution this is not true any more as one cannot fix the temperature of the dual field theory by choosing a value for $r_+$ alone, one has to also set a value for the non-conformal parameter $\eta$. Keeping this in mind, we provide the plot in the upper panel of Fig.\eqref{Butterfly}. Here, each point of the curve represents the value of the Lyapunov exponent for a particular temperature of the dual field theory which holographically corresponds to a particular value of the non-conformal parameter $\eta$, with the choice $r_+=1$ being taken without any loss of generality. It is also to be mentioned that the qualitative nature of the curve would remain same for other values of $r_+$ as well.
From Fig.\eqref{Butterfly} we observe that an increase in the value of the parameter $\eta$ decreases the value of the butterfly velocity $v_B$ and finally for $\eta=\sqrt{\frac{8d}{d-1}}$ it vanishes. This implies that due to the presence of non-conformality, the speed of information spreading ($v_B$) in the system gets decreased representing the delay in the growth of the initial perturbation provided to the system. For $\eta=\sqrt{\frac{8d}{d-1}}$, the system attains something known as the Lyapunov stability as for this particular value of $\eta$, the Hawking temperature of the non-conformal black brane is zero which in turn gives us a vanishing Lyapunov exponent, that is, $\lambda_L=0$. This can also be understood as the steady state. In this case, the signature of chaos in the system vanishes and the system becomes conservative. As a consequence of the Lyapunov stability, the butterfly effect also vanishes which is being manifested here as $v_B=0$ \cite{Ullmo2014}. We have already mentioned that this particular value of $\eta=\sqrt{\frac{8d}{d-1}}$ has the interpretation of being the upper bound (Gubser bound) of $\eta$ \cite{Gubser:2000nd,Gouteraux:2011ce}. The physics behind the Gubser bound is the following. For the Einstein-dilaton black brane geometry to be thermodynamically stable, $\eta^2$ should always be less than $\frac{8d}{d-1}$. This value is known as the Gubser bound. When $\eta^2$ exceeds this value, the Einstein-dilaton black brane and its dual theory are unstable thermodynamically \cite{Kulkarni:2012re,Park:2012lzs}. It can also be seen from the solution for $f(r)$ (eq.\eqref{EdBB}) that $c$ becomes negative when $\eta^2$ exceeds the Gubser bound. So $\eta^2\leq \frac{8d}{d-1}$ for $c$ to be positive. From our obtained results, we also confirm this observation from the point of view of chaos, as for $\eta<\sqrt{\frac{8d}{d-1}}$, the system is chaotic as both $\lambda_L$ and $v_B$ are positive and for $\eta=\sqrt{\frac{8d}{d-1}}$, the system is conservative. One can also recast the expressions given in eq.\eqref{result1} to the following forms
\begin{eqnarray}\label{NCcorrection}
	\lambda_L&=&\lambda_L^{(c)}\left(\frac{\beta^{(c)}}{\beta}\right)\nonumber\\
	v_B&=&v_B^{(c)}\sqrt{\left[1-\frac{(d-1)\eta^2}{8d}\right]}~.
\end{eqnarray}
The above forms helps us to point out the non-conformal corrections to the conformal results. Here, $\lambda_L^{(c)}$, $\beta^{(c)}$ and $v_B^{(c)}=\sqrt{\frac{d}{2(d-1)}}$ corresponds to the Lyapunov exponent, inverse Hawking temperature and butterfly velocity for the SAdS$_{d+1}$ black brane. As one shall have the SAdS$_{d+1}$ black brane solution in the conformal limit $\eta\rightarrow 0$, we denote the corresponding results as conformal results (that is why we use the superscript $(c)$).
\section{Two-sided holographic thermo mutual information (HTMI) and entanglement velocity}\label{section3}
In this section, we compute the HTMI between the two decoupled quantum mechanical systems existing on both left and right asymptotic boundaries of the two-sided black hole geometry \cite{Morrison:2012iz}. Further, we do this for both unperturbed and perturbed geometries in order to quantify the effect of shock wave on HTMI.
\subsection{HTMI for the unperturbed Einstein-dilaton black brane geometry}
 First, we holographically compute the two-sided thermo mutual information in absence of the shock wave. In order to do this, we consider two strip-like, identical subsystems of length $l$, namely, $A$ and $B$ which belongs to the left and right asymptotic boundaries respectively. The geometry of these strip-like subsystems ($A$ and $B$) can be specified as $-\frac{l}{2}\leq x_1 \leq \frac{l}{2}$ and $-\frac{L}{2}\leq x_m \leq \frac{L}{2}$ for $m=2,..,d-1$. As mentioned previously, the expression for MI is given by eq.\eqref{eq5}. To calculate this, one needs to compute the von Neumann entropies associated to subsystem $A$, $B$ and $A\cup B$. In the gauge/gravity set up, one can holographically do this by incorporating the RT proposal \cite{Ryu:2006bv,Ryu:2006ef,Hubeny:2007xt} which states that the von Neumann entropy of a subsystem (for example $A$) can be computed with the help of the extremal surface with minimal area $\gamma_A$. By incorporating this proposal, one can obtain the following result \cite{Saha:2020fon}
\begin{eqnarray}\label{SASB}
S_{\mathrm{vn}}(A)&=&S_{\mathrm{vn}}(B)\nonumber\\
&=&\frac{2L^{d-2}}{4}\int_{r_t}^{\infty} \frac{r^{p(d-3)}dr}{\sqrt{f(r)}\sqrt{1-\left(\frac{r_t}{r}\right)^{2p(d-1)}}}
\end{eqnarray}
where $r_t$ is the turning point of the static minimal surface of interest. On the other hand, the relation between the subsystem size $l$ and the turning point $r_t$ stands to be
\begin{eqnarray}\label{length}
l=2\int_{r_t}^{\infty} \frac{dr}{r^{2p}\sqrt{f(r)}\sqrt{\left(\frac{r}{r_t}\right)^{2p}-1}}~.
\end{eqnarray}
\begin{figure}[!h]
	\centering
	\includegraphics[width=0.5\textwidth]{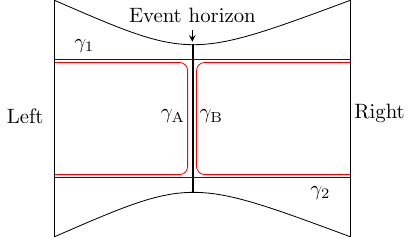}
	\caption{Schematic representation of the two-sided eternal black hole geometry in absence of the shock wave. Here, $\gamma_A$ and $\gamma_B$ (in red) are Ryu-Takayanagi surfaces associated to the subsystems $A$ and $B$. $\gamma_1$ and $\gamma_2$ (in blue) leads to the extremal surface $\gamma_{\mathrm{wormhole}}=\gamma_1\cup\gamma_2$ which connects the two decoupled boundaries (right and left).}
	\label{Wormhole1}
\end{figure}
The computation of $S_{\mathrm{vn}}(A\cup B)$ is tricky. For this, one proceeds with the surface $\gamma_{\mathrm{wormhole}}=\gamma_1 \cup \gamma_2$ which bifurcates the event horizon in order to connect the asymptotic boundaries of the two-sided eternal black hole spacetime. This is basically a non-traversable wormhole geometry which induces entanglement between the two decoupled theories (at right and left asymptotic boundaries of the spacetime geometry). It is to be noted that $\gamma_1$ corresponds to the $x_1=-\frac{l}{2}$ hyperplane and $\gamma_2$ corresponds to the $x_1=\frac{l}{2}$ hyperplane. If we consider the area of a single surface (with such properties), then symmetry tells us that total area will be four times of the area of the single surface. A graphical representation of the set up has been shown in Fig.\eqref{Wormhole1}. This leads to the following expression
\begin{eqnarray}\label{SWH}
	S_{\mathrm{vn}}(A\cup B; \alpha=0)&\equiv&\frac{\mathrm{Area}(\gamma_{\mathrm{wormhole}})}{4}\nonumber\\
	&=& \frac{4L^{d-2}}{4}\int_{r_+}^{\infty}\frac{r^{p(d-3)}dr}{\sqrt{f(r)}}~.
\end{eqnarray}\\
We now substitute the expressions from eq.\eqref{SASB} and eq.\eqref{SWH} in eq.\eqref{eq5} and obtain the following form of the two-sided HTMI (in the absence of shock wave)
\begin{eqnarray}\label{pureMI}
I(A:B;\alpha=0)&=&L^{d-2}\bigg[\int_{r_t}^{\infty} \frac{r^{p(d-3)}dr}{\sqrt{f(r)}\sqrt{1-\left(\frac{r_t}{r}\right)^{2p(d-1)}}}\nonumber\\
&&-\int_{r_+}^{\infty}\frac{r^{p(d-3)}dr}{\sqrt{f(r)}}\bigg]~.
\end{eqnarray}
The above expression represents the mutual correlation between the two decoupled theories living at the right and left asymptotic boundaries of the two-sided non-conformal black brane geometry (given in eq.\eqref{EdBB}). It can be seen that the expression is written in terms of the bulk coordinate $r$ and one can represent it in terms of the boundary coordinate (subsystem size $l$) with the help of eq.\eqref{length}.
\subsection{HTMI for the shock wave geometry}
\begin{figure}[!h]
	\centering
	\includegraphics[width=0.5\textwidth]{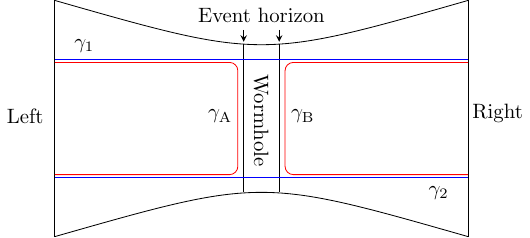}
	\caption{Schematic representation of the shock wave geometry. In presence of the shock wave, only the extremal surface $\gamma^{SW}_{\mathrm{wormhole}}=\gamma_1\cup\gamma_2$ (in blue) is getting affected.}
	\label{Wormhole2}
\end{figure}

\noindent We now proceed to compute the expression of $I(A:B)$ in presence of the shock wave modification of the original geometry. In order to do this, we follow the computational procedure shown in \cite{Leichenauer:2014nxa}. In this work, we are considering a homogeneous shock wave which is introducing deformation on the horizon along the $v$-coordinate which in turn stretches the wormhole geometry. We denote this new elongated wormhole geometry as $\gamma^{SW}_{\mathrm{wormhole}}=\gamma_1 \cup \gamma_2$. This leads to the fact that only the term $S_{\mathrm{vn}}(A:B)$ gets affected due to the presence of the shock wave as it is the only term in eq.\eqref{eq5} which depends on the wormhole geometry. A nice review related to this topic can be found in \cite{Kundu:2021nwp}. $S_{\mathrm{vn}}(A)$ and $S_{\mathrm{vn}}(B)$ remains unchanged as their associated extremal surfaces do not bifurcate the event horizon \cite{Hubeny:2012ry}. This can be graphically represented by a schematic diagram which has been provided in Fig.\eqref{Wormhole2}.
\noindent Keeping this in mind, for the shock wave geometry, one can write down the following expression
\begin{widetext}
\begin{eqnarray}\label{ShockI1}
I(A:B;\alpha)&=&S_{\mathrm{vn}}(A)+S_{\mathrm{vn}}(B)-S_{\mathrm{vn}}(A\cup B;\alpha)\nonumber\\
&=&S_{\mathrm{vn}}(A)+S_{\mathrm{vn}}(B)-S_{\mathrm{vn}}(A\cup B;\alpha=0)-\left[S_{\mathrm{vn}}(A\cup B;\alpha)-S_{\mathrm{vn}}(A\cup B;\alpha=0)\right]\nonumber\\
&=&I(A:B;\alpha=0)-S^{\mathrm{reg}}_{\mathrm{vn}}(A\cup B;\alpha)
\end{eqnarray}
where $I(A:B;\alpha=0)$ is given in eq.\eqref{pureMI} and $S^{\mathrm{reg}}_{\mathrm{vn}}(A\cup B;\alpha)$ represents a regularized expression for the von Neumann entropy which is free of the universal divergence term.
\end{widetext}
Furthermore, in the above computation we have also introduced $S_{\mathrm{vn}}(A\cup B;\alpha=0)$, which we have already computed in eq.\eqref{SWH}. The expression given in eq.\eqref{ShockI1} in turn means that we need to compute only the term $S^{\mathrm{reg}}_{\mathrm{vn}}(A\cup B;\alpha)$. In order to compute $S_{\mathrm{vn}}(A\cup B;\alpha)$ we first specify the parametrization of the corresponding HRT surfaces $r=r(t),x_1=\pm \frac{l}{2}$, and $-\frac{L}{2}\leq x_m \leq \frac{L}{2}$ for $m=2,..,d-1$. This leads to the corresponding area functional
\begin{eqnarray}
\mathrm{Area}(\gamma^{SW}_{\mathrm{wormhole}})=L^{d-2}\int r^{p(d-1)}\sqrt{-f(r)+\frac{\dot{r}^2}{r^{4p}f(r)}}dt~.\nonumber\\
\end{eqnarray}
It is to be noted that the above area functional for the wormhole has been computed by using the original geometry. However, unlike the unperturbed scenario (given in eq.\eqref{SWH}), one has to make use of the HRT proposal for the computation as the time-dependent shock wave perturbation makes the bulk direction time-dependent, that is, $r=r(t)$, and introduces time-dependent modification to the wormhole geometry \cite{Leichenauer:2014nxa} (see Fig.\eqref{Wormhole2})
\begin{eqnarray}
	\gamma_{\mathrm{wormhole}}~~\underrightarrow{\alpha(t,x_1)}~~\gamma^{SW}_{\mathrm{wormhole}}~.
\end{eqnarray} 
\noindent The Lagrangian density can be read off from the above action and has the form
\begin{eqnarray}
\mathcal{L}(r,\dot{r},t)=L^{d-2}r^{p(d-1)}\sqrt{-f(r)+\frac{\dot{r}^2}{r^{4p}f(r)}}~.
\end{eqnarray}
The corresponding Hamiltonian density can easily be obtained from the definition
\begin{eqnarray}
	\mathcal{H}(r,p_r,t)=p_r\dot{r}-\mathcal{L};~p_r=\frac{\partial \mathcal{L}}{\partial \dot{r}}~.
\end{eqnarray}
This in turn gives
\begin{eqnarray}
	\mathcal{H}(r,p_r)=r^{p(d-1)}\frac{f(r)}{\sqrt{-f(r)+\frac{\dot{r}^2}{r^{4p}f(r)}}}
\end{eqnarray}
where it can be observed that the Hamiltonian $\mathcal{H}(r,p_r)$ does not have an explicit dependency on time $t$. Hence, the Hamiltonian is a constant of motion. This can be written down as
\begin{eqnarray}
	\mathcal{H}(r,p_r) = \mathcal{C} \equiv \mathrm{constant}  
\end{eqnarray}
where the conserved quantity $\mathcal{C}$ can be evaluated at the point $r=r_0$ where $\dot{r}|_{r=r_0}=0$. This reads
\begin{eqnarray}
\mathcal{C}= \mathcal{H}\bigg|_{r=r_0}\equiv-r_0^{p(d-1)}\sqrt{-f(r_0)}~.
\end{eqnarray} 
The on-shell area functional therefore reads
\begin{eqnarray}
\mathrm{Area}(\gamma^{SW}_{\mathrm{wormhole}})=L^{d-1}\int\frac{r^{p(d-3)}dr}{\sqrt{f(r)-\left(\frac{r_0}{r}\right)^{2p(d-1)}f(r_0)}}~.\nonumber\\
\end{eqnarray}\\
On the other hand, the time-coordinate can be represented in the following way
\begin{eqnarray}
t(r)=\pm\int\frac{dr}{r^{2p}f(r)\sqrt{1-\left(\frac{r}{r_0}\right)^{2p(d-1)}\frac{f(r)}{f(r_0)}}}~.
\end{eqnarray}
We now proceed to specify the domain of integration for the above expressions. The domain of interest can be divided into three segments which has been pointed out in Fig.\eqref{SWRT}. It can be observed that segment II and segment III has the same area. Keeping these observations in mind, we write down the following form
\begin{widetext}
\begin{eqnarray}\label{New_S}
S_{\mathrm{vn}}(A\cup B;\alpha) = \frac{4L^{d-2}}{4}\left[\int_{r_+}^{\infty}\frac{r^{p(d-3)}dr}{\sqrt{f(r)-\left(\frac{r_0}{r}\right)^{2p(d-1)}f(r_0)}}+2\int_{r_0}^{r_+}\frac{r^{p(d-3)}dr}{\sqrt{f(r)-\left(\frac{r_0}{r}\right)^{2p(d-1)}f(r_0)}}\right]~.
\end{eqnarray}
\begin{figure}[!h]
	\centering
	\includegraphics[width=0.4\textwidth]{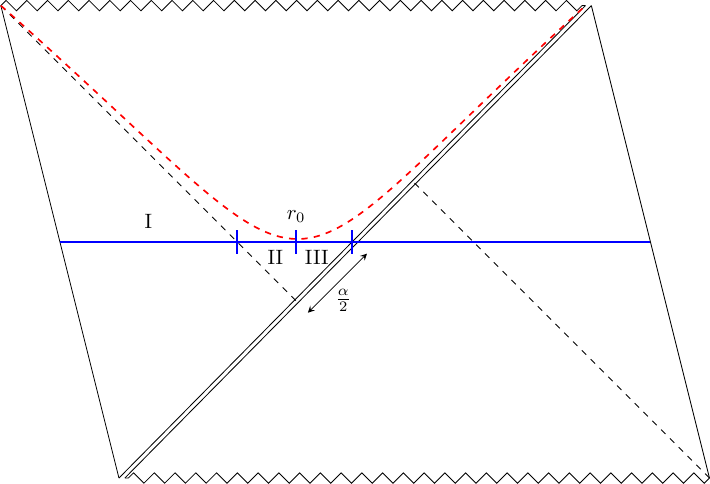}
	\caption{Deformation of the extremal surface $\gamma^{SW}_{\mathrm{wormhole}}$ (in blue) due to the shock wave. We have divided the left half of the surface into three parts segment $\mathrm{I}$ which spans from the boundary ($r=\infty$) to the horizon $r_+$ and both segment $\mathrm{II}$ and segment $\mathrm{III}$ spans from $r_0$ to the horizon $r_+$. Thus, segment $\mathrm{II}$ and segment $\mathrm{III}$ have same area.}
	\label{SWRT}
\end{figure}	
\end{widetext}
With the above result in hand, we now make use of the expression for $S_{\mathrm{vn}}(A\cup B;\alpha=0)$ (given in eq.\eqref{SWH}) in order to obtain the regularized version of $S_{\mathrm{vn}}(A\cup B;\alpha)$. 
\begin{widetext}
\noindent This reads
\begin{eqnarray}\label{S_New}
S^{\mathrm{reg}}_{\mathrm{vn}}(A\cup B;\alpha)&=&S_{\mathrm{vn}}(A\cup B;\alpha)-S_{\mathrm{vn}}(A\cup B;\alpha=0)\nonumber\\
&=&\frac{4L^{d-2}}{4}\Bigg[\int_{r_+}^{\infty}r^{p(d-3)}\left(\frac{1}{\sqrt{f(r)-\left(\frac{r_0}{r}\right)^{2p(d-1)}f(r_0)}}-\frac{1}{\sqrt{f(r)}}\right)dr+2\int_{r_0}^{r_+}\frac{r^{p(d-3)}dr}{\sqrt{f(r)-\left(\frac{r_0}{r}\right)^{2p(d-1)}f(r_0)}}\Bigg]~.\nonumber\\
\end{eqnarray}	
\end{widetext}	
As we have explained in eq.\eqref{ShockI1}, the above expression of $S^{\mathrm{reg}}_{\mathrm{vn}}(A\cup B;\alpha)$ along with the expression of $I(A:B;\alpha=0)$ (given in eq.\eqref{pureMI}) leads us to the desired result of $I(A:B;\alpha)$. It is to be observed from the above expression of $S^{\mathrm{reg}}_{\mathrm{vn}}(A\cup B;\alpha)$ that it is a function of $r_0$. Before we proceed further we would like to make few comments. From eq.\eqref{S_New} one can observe that in the limit $r_0\rightarrow r_+$, $S_{\mathrm{vn}}(A\cup B;\alpha)=S_{\mathrm{vn}}(A\cup B;\alpha=0)$. We would like to point out that the emergence of the point $r_0$ in the black hole interior (see Fig.\eqref{SWRT}) is precisely due to the shock wave perturbation of the original geometry. This in turn means that we need to find the relation between $\alpha$ (the shock wave parameter) and $r_0$ so that we can depict the variation of $S^{\mathrm{reg}}_{\mathrm{vn}}(A\cup B;\alpha)$ with respect to the shock wave parameter. In order to do this, we follow the approach given in \cite{Leichenauer:2014nxa}. As we have mentioned earlier, our domain of interest can be divided into three segments, namely, $\mathrm{I}$, $\mathrm{II}$ and $\mathrm{III}$. It is precisely in the partitioning of the concerned region into three segments that the information the shock wave has entered. In terms of the Kruskal coordinates, the segment $\mathrm{I}$ connects $(u,v)=(1,-1)$ (boundary) to $(u,v)=(u_1,0)$ (horizon). Segment $\mathrm{II}$ connects $(u,v)=(u_1,0)$ (horizon) to the point $r_0$ which is at $(u,v)=(u_2,v_2)$ and segment $\mathrm{III}$ connects the point $r_0$, that is, $(u,v)=(u_2,v_2)$ to $(u,v)=(0,\alpha/2)$. Keeping these coordinates in mind and by using the Kruskal coordinates, one can show the following relations
\begin{widetext}
\begin{eqnarray}\label{relationuv}
	u_1^2&=&\exp(\frac{4\pi}{\beta}\int_{r_+}^{\infty}\frac{dr}{r^{2p}f(r)}\left[\frac{1}{\sqrt{1-\left(\frac{r}{r_0}\right)^{2p(d-1)}\frac{f(r)}{f(r_0)}}}-1\right]),\nonumber\\
	u_2^2&=&\exp(\frac{4\pi}{\beta}\int_{r_0}^{r_+}\frac{dr}{r^{2p}f(r)}\left[\frac{1}{\sqrt{1-\left(\frac{r}{r_0}\right)^{2p(d-1)}\frac{f(r)}{f(r_0)}}}-1\right]),\nonumber\\
	v_2&=&\frac{1}{u_2}\exp(-\frac{4\pi}{\beta}\int_{r_0}^{\bar{r}}\frac{dr}{r^{2p}f(r)})~.
\end{eqnarray} 	
\end{widetext}
In the last of the above equations, $\bar{r}$ point resides inside the horizon at which $r_*=0$. From segment $\mathrm{III}$, one can show the following relation by incorporating the variation in the $v$-coordinate
\begin{widetext}
\begin{eqnarray}
	\frac{\alpha^2}{4v_2^2}=\exp(\frac{4\pi}{\beta}\int_{r_+}^{r_0}\frac{dr}{r^{2p}f(r)}\left[\frac{1}{\sqrt{1-\left(\frac{r}{r_0}\right)^{2p(d-1)}\frac{f(r)}{f(r_0)}}}-1\right])=\frac{u_1^2}{u_2^2}~.
\end{eqnarray}
Now by using the relation given by eq.\eqref{relationuv} in the above equation and after some simplifications, one obtains the following relation
\begin{eqnarray}\label{SWr0}
\alpha(r_0)=2\exp(\xi_I+\xi_{II}+\xi_{III})
\end{eqnarray}
where 
\begin{eqnarray}
\xi_I&=&\frac{4\pi}{\beta}\int_{\bar{r}}^{r_0}\frac{dr}{r^{2p}f(r)}\nonumber\\
\xi_{II}&=&\frac{2\pi}{\beta}\int_{r_+}^{\infty}\frac{dr}{r^{2p}f(r)}\left[1-\frac{1}{\sqrt{1-\left(\frac{r}{r_0}\right)^{2p(d-1)}\frac{f(r)}{f(r_0)}}}\right]\nonumber\\
\xi_{III}&=&\frac{4\pi}{\beta}\int_{r_0}^{r_+}\frac{dr}{r^{2p}f(r)}\left[1-\frac{1}{\sqrt{1-\left(\frac{r}{r_0}\right)^{2p(d-1)}\frac{f(r)}{f(r_0)}}}\right]~.\nonumber\\
\end{eqnarray}
\end{widetext}

\begin{figure}[!h]
	\centering
	\includegraphics[width=0.46\textwidth]{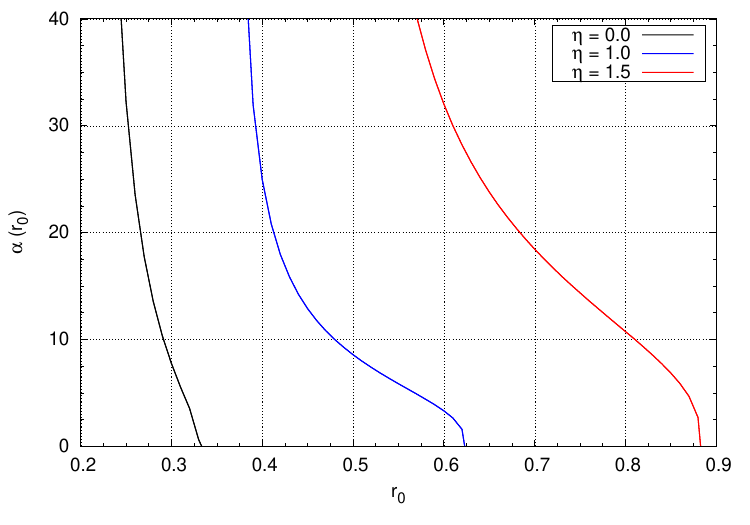}
	\caption{Behaviour of the shock wave parameter $\alpha(r_0)$ with respect to the turning point $r_0$. We have set $\beta=4\pi$ and $d=3$. Here, the right curve is for $\eta=1.5$, the middle curve is for $\eta=1.0$ and the left curve is for $\eta=0.0$.}
	\label{SWprofile}
\end{figure}
\begin{figure}[!h]
	\centering
	\includegraphics[width=0.47\textwidth]{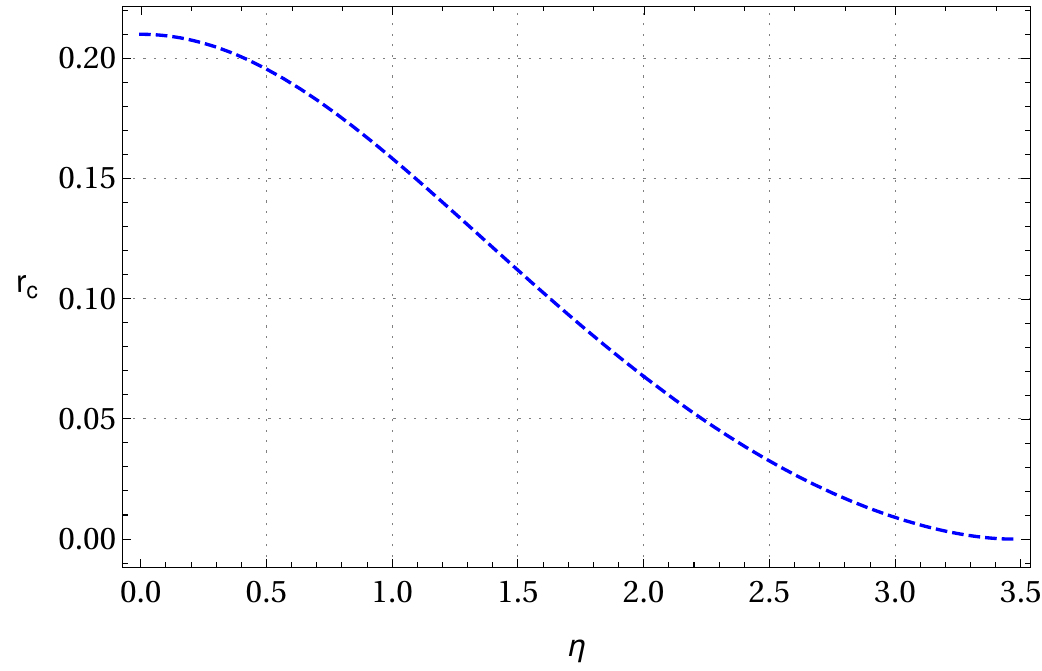}
	\caption{Effect of the non-conformal parameter $\eta$ on $r_c$. We have set $d=3$ and $\beta=4\pi$.}
	\label{Divergence}
\end{figure}

In Fig.\eqref{SWprofile}, we have graphically represented the relation given in eq.\eqref{SWr0} with $\beta=4\pi$. We observe that $\alpha(r_0)$ decreases with the increase in the value of the $r_0$ and ultimately vanishes at $r_0=r_+$. The lesson that one can learn from this observation is the following. For a fixed $\beta$, an increase in the value of the non-conformal parameter $\eta$ increases the value of $r_+$. This can be understood from the relation provided in eq.\eqref{EHinBeta}. This in turn means that the presence of non-conformality leads to a higher value of $r_0$ at which the shock wave parameter vanishes.  It is also to be noted that for a critical value of $r_0$, namely, at $r_0=r_c$ the shock wave parameter $\alpha(r_0)$ diverges. In fact, it can also be observed that at this particular value $r_0=r_c$, only $\xi_{III}$ diverges which depicts the fact that in the limit $r_0\rightarrow r_c$, $\xi_{III}$ is the dominating piece in the expression of $\alpha(r_0)$ (given in eq.\eqref{SWr0}.) One can derive the value of $r_c$ by performing a Taylor expansion of the integrand of $\xi_{III}$ around $r_0\approx r_c$ and equating the coefficient of $(r_0-r_c)$ to zero \cite{Jahnke:2017iwi,Fischler:2018kwt,Avila:2018sqf}. We shall follow this approach to derive the explicit expression of $r_c$. Firstly, the expression of $\xi_{III}$ around $r\approx r_0$ has the following form
\begin{widetext}
\begin{eqnarray}\label{EqNew2}
\xi_{III}&\approx&\frac{4\pi}{\beta}\int_{r_0}^{r_+}\frac{dr}{r_0^{2p}f(r_0)}\left[1-\frac{1}{\sqrt{\left(-\frac{2p(d-1)}{r_0}-\frac{f^{\prime}(r_0)}{f(r_0)}\right)(r-r_0)}}\right]~.
\end{eqnarray}	
\end{widetext}
The above expression diverges at $$-\frac{2p(d-1)}{r_0}-\frac{f^{\prime}(r_0)}{f(r_0)}|_{r_0=r_c}=0~.$$ By solving this one obtains
\begin{eqnarray}\label{critr}
r_c=r_+\left[1-\frac{c}{2p(d-1)}\right]^{\frac{1}{c}}~.
\end{eqnarray}
In the limit $\eta\rightarrow 0$, one obtains the conformal result \cite{Leichenauer:2014nxa}
\begin{eqnarray}
r_c=r_+\left[\frac{d-2}{2(d-1)}\right]^{\frac{1}{d}}~.	
\end{eqnarray}
We now make use of the relation given in eq.\eqref{EHinBeta} in order to recast the expression of $r_c$ (given in eq.\eqref{critr}) to the following form
	\begin{eqnarray}
	r_c = \left(\frac{\beta c}{4\pi}\right)^{\frac{1}{1-2p}}\left[1-\frac{c}{2p(d-1)}\right]^{\frac{1}{c}}~.
	\end{eqnarray}
In Fig.\eqref{Divergence}, we have graphically represented the above expression of $r_c$ for a fixed value of $\beta$ in order to capture the effect of non-conformality on it. We observe that non-conformality decreases the value of $r_c$ which implies the fact that non-conformality helps to probe the black hole interior further. In Fig.\eqref{EEMI}, the behaviour of $S^{\mathrm{reg}}_{\mathrm{vn}}(A\cup B;\alpha)$ and $\frac{I(A:B;\alpha)}{I(A:B;\alpha=0)}=1-\frac{S^{\mathrm{reg}}_{\mathrm{vn}}(A\cup B;\alpha)}{I(A:B;\alpha=0)}$ with respect to the logarithm of the shock wave parameter ($\log\alpha$) has been provided. We observe that for a fixed value of $\log\alpha$, $S^{\mathrm{reg}}_{\mathrm{vn}}(A\cup B;\alpha)$ increases with the increase in the value of the non-conformal parameter $\eta$. On the other hand, due to non-cormality, $S^{\mathrm{reg}}_{\mathrm{vn}}(A\cup B;\alpha)$ becomes equal to $I(A:B;\alpha=0)$ (resulting in $\frac{I(A:B;\alpha)}{I(A:B;\alpha=0)}=0$) for a smaller value of $\log\alpha$.
\begin{widetext}
	~\\
	\begin{figure}[!htb]
		\centering
		\begin{minipage}[t]{0.48\textwidth}
			\centering
			\includegraphics[width=\textwidth]{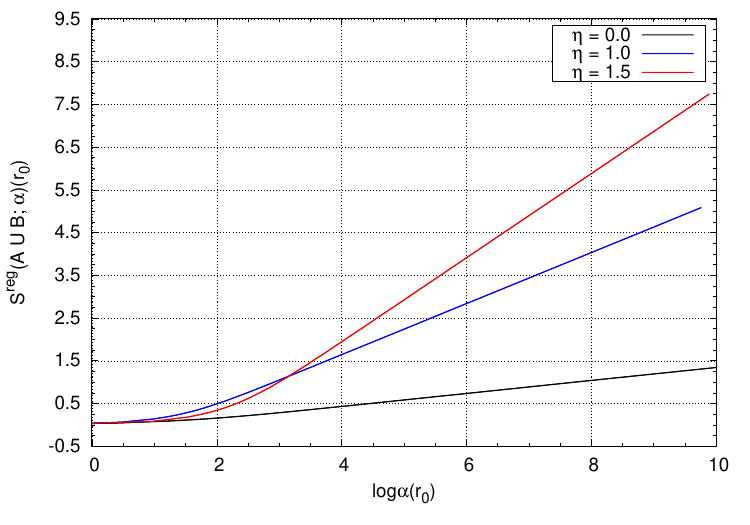}\\
			{\footnotesize  The upper curve is for $\eta=1.5$, the middle curve is for $\eta=1.0$ and the lower curve is for $\eta=0.0$.}
		\end{minipage}\hfill
		\begin{minipage}[t]{0.48\textwidth}
			\centering
			\includegraphics[width=\textwidth]{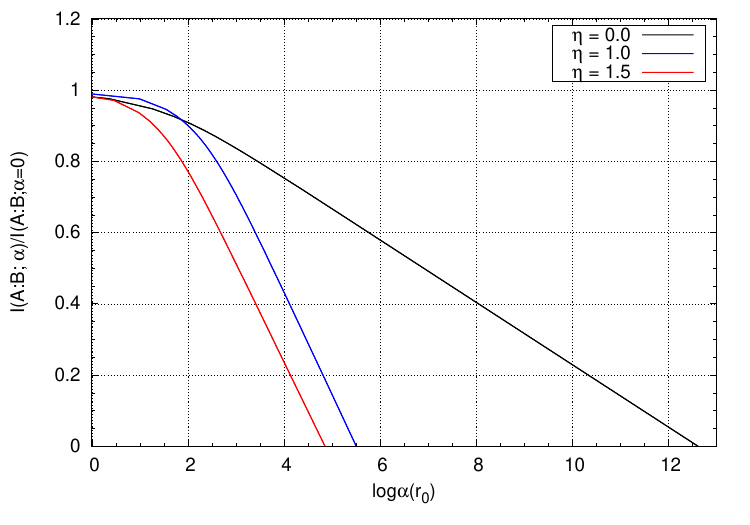}\\
			{\footnotesize  The upper curve is for $\eta=0.0$, the middle curve is for $\eta=1.0$ and the lower curve is for $\eta=1.5$.}
		\end{minipage}
		\caption{The left plot represents the behaviour of $S^{\mathrm{reg}}_{\mathrm{vn}}(A\cup B;\alpha)$ with respect to $\log\alpha$. The right plot captures the behaviour of the thermo-mutual information in presence of the shock wave. Here, we have set $\beta=4\pi$, $L=1$ and $d=3$.}
		\label{EEMI}
	\end{figure}	
\end{widetext}
\newpage
\subsection{Entanglement velocity}
\noindent We now proceed to study the behaviour of $S^{\mathrm{reg}}_{\mathrm{vn}}(A\cup B;\alpha)$ with respect to the time $t$ at which the initial perturbation was added. It can be observed that $S^{\mathrm{reg}}_{\mathrm{vn}}(A\cup B;\alpha)$ grows linearly with respect to $\log\alpha$ which in turn means it grows linearly with respect to $t$ (as $\alpha \approx e^{\frac{2\pi}{\beta}t}$). This can be observed from the left plot of Fig.\eqref{EEMI}. This linear behaviour of $S^{\mathrm{reg}}_{\mathrm{vn}}(A\cup B;\alpha)$ in turn helps us to quantify the spreading of entanglement in a chaotic system by introducing the entanglement velocity in this set up. In order to capture the behaviour of $S^{\mathrm{reg}}_{\mathrm{vn}}(A\cup B;\alpha)$ around $r_0\approx r_c$, we first expand $S^{\mathrm{reg}}_{\mathrm{vn}}(A\cup B;\alpha)$ upto linear order in $(r-r_0)$. This leads to the following form
\begin{widetext}
\begin{eqnarray}
S^{\mathrm{reg}}_{\mathrm{vn}}(A\cup B;\alpha)&\approx& 2L^{d-2}\sqrt{-f(r_0)} r_0^{p(d-1)}\int_{r_0}^{r_+}\frac{dr}{r_0^{2p}f(r_0)\sqrt{\left(-\frac{2p(d-1)}{r_0}-\frac{f^{\prime}(r_0)}{f(r_0)}\right)(r-r_0)}}\nonumber\\
&=&2L^{d-2}\sqrt{-f(r_0)} r_0^{p(d-1)}\int_{r_0}^{r_+} \frac{dr}{r_0^{2p}f(r_0)}\left[1-\frac{1}{\sqrt{\left(-\frac{2p(d-1)}{r_0}-\frac{f^{\prime}(r_0)}{f(r_0)}\right)(r-r_0)}}\right]\nonumber\\
&-&2L^{d-2}\sqrt{-f(r_0)} r_0^{p(d-1)}\int_{r_0}^{r_+}\frac{dr}{r_0^{2p}f(r_0)}~.
\end{eqnarray}	
\end{widetext}
Now, by using the relation given in eq.\eqref{EqNew2}, we can write down the following form of $S^{\mathrm{reg}}_{\mathrm{vn}}(A\cup B;\alpha)$ and proceed to consider the $r_0\rightarrow r_c$ limit
\begin{eqnarray}
S^{\mathrm{reg}}_{\mathrm{vn}}(A\cup B;\alpha)&\approx& 2L^{d-2}\sqrt{-f(r_0)} r_0^{p(d-1)}\nonumber\\
&&\times\left(\frac{\beta}{4\pi}\right)\log\alpha(r_0)\bigg|_{r_0=r_c}~.
\end{eqnarray}
Keeping in mind the exponential growth of the given perturbation, that is, $\alpha\sim \exp^{\frac{2\pi t}{\beta}}$, one can write down the following equation \cite{Hartman:2013qma,Liu:2013iza,Liu:2013qca,Mezei:2016wfz,Mezei:2016zxg}
\begin{eqnarray}
\frac{dS^{\mathrm{reg}}_{\mathrm{vn}}(A\cup B;\alpha)}{dt}&=&2L^{d-2}\sqrt{-f(r_c)}r_c^{p(d-1)}\left(\frac{\beta}{4\pi}\right)\left(\frac{2\pi}{\beta}\right)\nonumber\\
&=&4L^{d-2}\sqrt{-f(r_c)}\left(\frac{r_c}{r_+}\right)^{p(d-1)}\left(\frac{r_+^{p(d-1)}}{4}\right)\nonumber\\
&=&s_{th} \mathcal{A}_{\Sigma}v_{en}
\end{eqnarray}
where $s_{th}$ is the thermal entropy density $s_{th}=\frac{r_+^{p(d-1)}}{4}$, $\mathcal{A}_{\Sigma}=4L^{d-2}$ is the area of the hyperplane and $v_{en}$ is the entanglement velocity which has the following form
\begin{eqnarray}
v_{en}=\left(\frac{r_c}{r_+}\right)^{p(d-1)}\sqrt{-f(r_c)}~.
\end{eqnarray}
By using the explicit value of $r_c$ (given in eq.\eqref{critr}), one can obtain an exact expression for the entanglement velocity. This reads
\begin{eqnarray}\label{entanglementV}
v_{en}=\left[1-\frac{c}{2p(d-1)}\right]^{\frac{p(d-1)}{c}}\sqrt{\frac{c}{2p(d-1)-c}}~.
\end{eqnarray}
\begin{figure}[!h]
		\centering
		\includegraphics[width=0.47\textwidth]{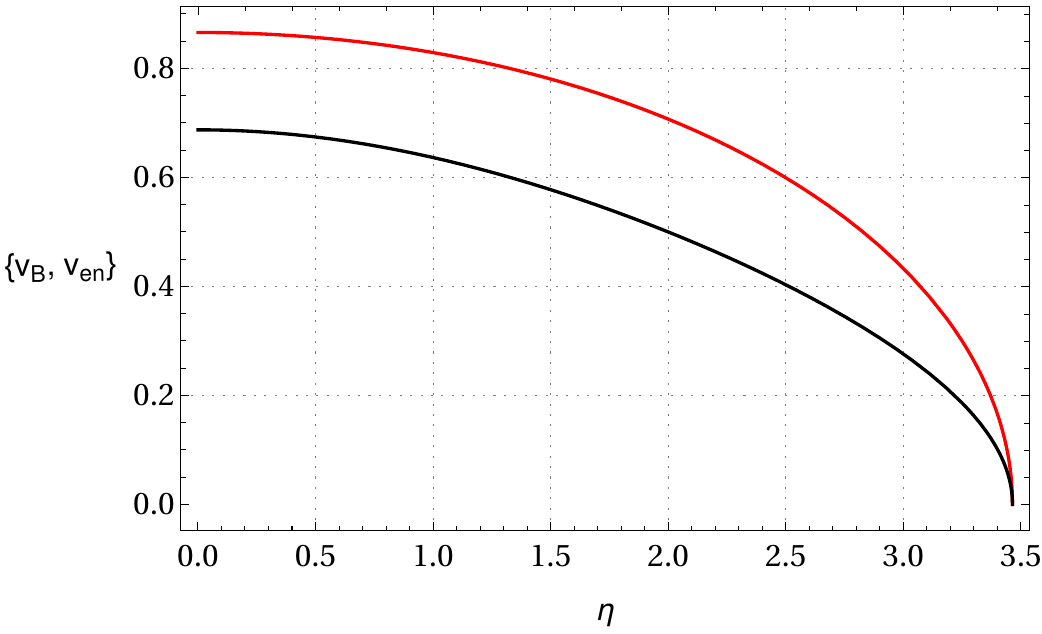}
		\caption{Comparison between the variations of the butterfly velocity (the upper curve) and entanglement velocity (the lower curve) with respect to the non-conformal parameter $\eta$.}
	\label{EV}
\end{figure}
In the limit $\eta\rightarrow0$, one obtains the standard conformal result (result for SAdS$_{d+1}$). This reads \cite{Hartman:2013qma}
\begin{eqnarray}
v_{en}=\frac{\sqrt{d}(d-2)^{\frac{1}{2}-\frac{1}{d}}}{[2(d-1)]^{1-\frac{1}{d}}}~.
\end{eqnarray}
From our obtained result of entanglement velocity (given in eq.\eqref{entanglementV}), we observe that similar to the butterfly velocity and Lyapunov exponent, it also decreases with the increase in the value of the non-conformal parameter $\eta$. Furthermore, $v_{en}$ also vanishes for $\eta=\sqrt{\frac{8d}{(d-1)}}$ representing the complete disruption of quantum entanglement. We have represented our observations graphically in Fig.\eqref{EV}. Fig.\eqref{EV} also reveals that in the presence of non-conformality, entanglement velocity still satisfies the property $v_{en}\leq v_B$. Further, both of these velocities are always less than the speed of light (as speed of light $v_c=1$).
\section{Pole-skipping analysis: Lyapunov exponent and Butterfly velocity}\label{section4}
\noindent In this section, we will point out the special points, known as the pole-skipping points, in the complex $(\omega,k)$ plane at which the near-horizon solution of the bulk field is ill-defined which leads to the non-unique nature of the corresponding retarded Green's function. As mentioned earlier, the location of these points in the complex $(\omega,k)$ (upper half or lower half) depends on the spin $s$ of the bulk field. As we have already shown in eq.\eqref{eq8}, only the upper half pole-skipping points are related to the Lyapunov exponent and the butterfly velocity. As we are interested in the near horizon analysis of the bulk field equation of motion, we recast the bulk metric (given in eq.\eqref{EdBB}) in the ingoing Eddington-Finkelstein coordinate system
\begin{eqnarray}
	v=t+r_*~.
\end{eqnarray}\\
In the above coordinate system, the bulk metric takes the following form
\begin{eqnarray}\label{EFform}
ds^2=-r^{2p}f(r)dv^2+2dvdr+h(r)\sum_{i=1}^{d-1}dx_i^2~,~h(r)=r^{2p}~.\nonumber\\
\end{eqnarray}
On the other hand, we assume the following linear perturbation of the metric and the dilaton field (in the $x_1$ direction)
\begin{eqnarray}
	\delta g_{\mu\nu}(v,r,x_1)&=& \delta g_{\mu\nu}(r)e^{-i\omega v+ikx_1}\nonumber\\
	\delta \phi(v,r,x_1)&=&\delta\phi(r)e^{-i\omega v+ikx_1}~.
\end{eqnarray}
In the sound mode, the relevant perturbations are $\delta g_{vv}$, $\delta g_{vr}$, $\delta g_{vx_1}$, $\delta g_{rr}$, $\delta g_{rx_1}$, $\delta g_{x_ix_i}$ along with $\delta\phi$ which can couple to any of these metric fluctuations. As the location of the pole-skipping point $(\omega_*,k_*)$ depends on the behaviour of the background metric on the horizon, one has to consider the near-horizon behaviour of these perturbations \cite{Blake:2018leo,Blake:2019otz,Blake:2021hjj}. In the near-horizon regime, the perturbations behave in the following way
\begin{eqnarray}
\delta g_{\mu\nu}(r) &=& \delta g_{\mu\nu}^{(0)}+\delta g_{\mu\nu}^{(1)}(r-r_+)+\mathcal{O}(r-r_+)^2...\nonumber\\
\delta\phi(r)&=&\delta\phi^{(0)}+\delta\phi^{(1)}(r-r_+)+\mathcal{O}(r-r_+)^2...~.
\end{eqnarray}
We now follow the approach given in \cite{Blake:2018leo}.
\begin{widetext}
	\noindent By substituting the above forms of perturbations in the Einstein field equations (given in eq.\eqref{EOM}) and by considering the near-horizon limit, one observes that the $vv$-component of the Einstein-field equation assumes the following universal form \cite{Blake:2018leo} 
	\begin{eqnarray}
	\left(-i\frac{(d-1)}{2}\omega h^{\prime}(r_+)+k^2\right)\delta g_{vv}^{(0)}-i(2\pi T_H+i\omega)\left(\omega\delta g_{x_ix_i}^{(0)}+2k\delta g_{vx_1}^{(0)}\right)=0~.
	\end{eqnarray}
	By incorporating the background metric informations from eq.\eqref{EFform}, we obtain the following form for the case we have in hand
	\begin{eqnarray}
	\left(-i\frac{(d-1)}{2}\omega\partial_r r^{2p}\vert_{r=r_+}+k^2\right)\delta g_{vv}^{(0)}-i(2\pi T_H+i\omega)\left(\omega\delta g_{x_ix_i}^{(0)}+2k\delta g_{vx_1}^{(0)}\right)=0~.
	\end{eqnarray}
\end{widetext}	
From the above equation, one can easily point out the special values of $\omega$ and $k$ for which the equation gets satisfied. The values of $\omega$ read
\begin{eqnarray}\label{specialomega}
	\omega=i2\pi T_H\equiv \omega_*~.
\end{eqnarray}
On the other hand, for $k$, the corresponding special value is given by
\begin{eqnarray}\label{specialk}
k=\sqrt{i\frac{(d-1)}{2}\omega_*\partial_r r^{2p}\vert_{r=r_+}}
=i\frac{2\pi T_H}{\sqrt{\frac{c}{2p(d-1)}}}
=k_*~.
\end{eqnarray}
We now make use of the relations given in eq.\eqref{eq8} along with the above obtained values of $\omega_*$ and $k_*$ to obtain the following results
\begin{eqnarray}
	\lambda_L&=&\frac{2\pi}{\beta}\\
	v_B&=&\sqrt{\frac{c}{2p(d-1)}}\equiv\frac{1}{4}\sqrt{\frac{8d}{d-1}-\eta^2}~.
\end{eqnarray}
The above results agrees perfectly with that obtained from the shock wave analysis (eq.\eqref{result1}). A few remarks are in order now. From the above computation, one can conclude that the metric perturbation in the sound channel leads to the energy density Green's functions which has the pole-skipping points in the upper half of the complex $(\omega,k)$ plane. The importance of these points lying in the upper half plane is that they lead to the parameters of chaos, based upon the relations given in eq.\eqref{eq8}. This further verifies the previously mentioned statement that for a strongly coupled theory with holographic dual which is maximally chaotic, the pole-skipping points are related to the parameters of chaos. However, it is to be kept in mind that the association of the pole-skipping points with the parameters of chaos can be made as long as the pole-skipping points are in the upper half of the complex $(w,k)$ plane. In the Appendix, we show that for a scalar field perturbation, the pole-skipping points lie in the lower half of the complex $(\omega,k)$ plane and hence are not related to the parameters of chaos even for a maximally chaotic system.
\section{Conclusion}\label{section5}
\noindent We now summarize our findings. In this work, we have holographically studied the behaviour of the parameters of chaos in presence of non-conformality. By incorporating the gauge/gravity framework, we have introduced the two-sided black hole geometry which is the well-known dual description for the thermofield doublet state. This realization helps us to quantify the effects of chaos on the correlation which exists between the right and left boundary theories of the two-sided geometry. The non-conformality in the boundary theories has been holographically introduced by considering the black brane solution of the Einstein-dilaton theory where the dilaton potential is of Liouville type. The asymptote of the black brane geometry is a warped geometry instead of a pure AdS geometry. This implies that the boundary theory is non-conformal as the scale transformations are broken, however, it is relativistic as the Poinc\'are symmetry is still restored.  The black brane solution is associated with a parameter $\eta$ which characterizes the deviation from conformality as in the limit $\eta \rightarrow0$, one obtains the usual Schwarzschild black brane solution. In order to keep things general, we have considered the gravitational theory is of $(d+1)$-dimensional which in turn means the boundary theories are $d$-dimensional. In order to compute the parameters of chaos which are Lyapunov exponent and the butterfly velocity, we obtain the shock wave geometry corresponding to the black brane solution under consideration. The shock wave geometry arises due to the introduction of a tiny pulse of energy in the geometry (or in dual sense, adding of perturbation at the boundary theory). Due to presence of event horizon, the energy of the pulse gets blue-shifted resulting in a non-trivial modification to the original geometry. The obtained results for the Lyapunov exponent and butterfly velocity from the shock wave analysis reveal some interesting observations. We observed that the form of the Lyapunov exponent remains unchanged, that is, $\lambda_L = 2\pi T_{H}$, as this form is universal for any maximally chaotic holographic theory. It is important to note that the Lyapunov exponent depends only on the Hawking temperature $T_H$ (which is also the temperature of the dual field theory).
%The Hawking temperature in general depends on the radius of the event horizon ($r_+$) and hence the Lyapunov exponent depends on $r_+$ only.  However, for the model that we have considered in this paper, the Lyapunov exponent not only depends on $r_+$ but also depends on the non-conformal parameter $\eta$.
On the other hand, the effect of non-conformality on the butterfly velocity ($v_B$) is pretty straight forward to see from the computed expression. Further, we observe that $v_B$ decreases with increase in the value of the non-conformal parameter $\eta$ and finally it vanishes for $\eta=\sqrt{\frac{8d}{d-1}}$. We believe this represents Lyapunov stability for the system as for this particular value of $\eta$, the Hawking temperature of the black brane becomes zero (irrespective of the value of $r_+$) which leads to the vanishing of the Lyapunov exponent, that is, $\lambda_L=0$. This particular value of $\eta$ matches perfectly with the previously known upper bound of $\eta$, known as the Gubser bound. Our results also confirm the value of this bound from the point of view of chaos. Our results also indicate that non-conformality helps to suppress the chaotic nature for a system.\\
On the other hand, it is a well-known fact that the left and right boundary theories of a two-sided geometry share a non-vanishing quantum correlation between them which can be characterized by the thermo mutual information (TMI) between the mentioned two sides. In order to observe the effects of chaos and non-conformality, we compute the holographic thermo mutual information (HTMI) both in presence and absence of the shock wave. We observe that non-conformality increases the existing entanglement between the boundary theories, namely, $A$ and $B$. This can be understood from the behaviour of $S^{\mathrm{reg}}_{\mathrm{vn}}(A\cup B;\alpha)$ as its value increases with the increase in the value of $\eta$ (for a fixed value of the shock wave parameter $\log\alpha$). We also note that there is critical value for $r_0$, namely, $r_c$ at which the shock wave parameter diverges. Furthermore, an increase in value of $\eta$ decreases the value of $r_c$. This in turn means that the presence of non-conformality helps us to probe further the black hole interior as the point $r_0$ resides inside the black hole interior. In order to understand the spreading of entanglement for a chaotic system, we proceed to compute the entanglement velocity. We observe that similar to the butterfly velocity, this quantity also decreases with the increase in non-conformality, maintaining the bound $v_B<v_{en}$. Finally, we once again obtain the Lyapunov exponent and butterfly velocity from the lowest pole-skipping points in the upper half of the complex-$(\omega,k)$ plane. We observe that the obtained results matches perfectly with that obtained from the shock wave analysis. We also compute higher order pole-skipping points (in the lower half of the complex-$(\omega,k)$ plane) by considering scalar field fluctuation in the bulk geometry and give the results in the Appendix.
\section*{Appendix: Scalar field perturbation and pole-skipping points in the lower half of the complex $(\omega,k)$ plane}
In this Appendix, we compute the pole-skipping points by considering scalar field fluctuations in the gravitational background. We start our analysis by considering a minimally coupled massive scalar field $\Psi(v,r,x_1)$, the equation of motion for which is governed by the following Klein-Gordon (KG) equation
\begin{eqnarray}\label{KG}
\frac{1}{\sqrt{-g}}\partial_{\mu}\left(\sqrt{-g}g^{\mu\nu}\partial_{\nu}\Psi\right)-m^2\Psi=0
\end{eqnarray}
with the background metric being given by eq.\eqref{EFform}.
\begin{widetext}
	\noindent The above equation assumes the following form in the Eddington-Finkelstein coordinate
	\begin{eqnarray}\label{KGF}
	&&p(d-1)r^{p(d-1)-1}\partial_v\Psi+2r^{p(d-1)}\partial_v\partial_r\Psi+p(d+1)r^{p(d+1)-1}f(r)\partial_r\Psi+r^{p(d+1)}f^{\prime}(r)\partial_r\Psi+r^{p(d+1)}f(r)\partial^2_r\Psi+r^{p(d-3)}\partial^2_{x_1}\Psi\nonumber\\
	&&-r^{p(d-1)}m^2\Psi=0~.
	\end{eqnarray}	
	\noindent We now consider the following Fourier decomposition of the massive scalar field
	\begin{eqnarray}\label{EqNew3}
	\Psi(v,r,x_1)=\psi(r) e^{-i\omega v+ikx_1}~.
	\end{eqnarray}
	By substituting eq.\eqref{EqNew3} in eq.\eqref{KGF}, we obtain 
	\begin{eqnarray}\label{KGF2}
	\psi^{\prime\prime}(r)+\Delta_1(r)\psi^{\prime}(r)+\Delta_2(r)\psi(r)=0
	\end{eqnarray}
	where
	\begin{eqnarray}
	\Delta_1(r)&=&\frac{r^{2p}f^{\prime}(r)+p(d+1)r^{2p-1}f(r)-2i\omega}{r^{2p}f(r)}\nonumber\\
	\Delta_2(r)&=&-\frac{\frac{k^2}{r^{2p}}+m^2+\frac{i\omega}{r}p(d-1)}{r^{2p}f(r)}~.
	\end{eqnarray}
\end{widetext}
In order to find out the pole-skipping points, one needs to consider the near-horizon limit. The near-horizon expansion for the scalar field reads $\psi(r)$
\begin{eqnarray}
\psi(r)=\psi_0+\psi_1(r-r_+)+\psi_2(r-r_+)^2+...~.
\end{eqnarray}
By substituting the above near-horizon form of $\psi(r)$ in eq.\eqref{KGF2} and by equating the coefficients of $(r-r_+)^j$ (for $j=0,1,2..$) to zero, one obtains a set of linear equations which read
\begin{eqnarray}
\chi_{00}\psi_0+\chi_{01}\psi_1&=&0,\nonumber\\
\chi_{10}\psi_0+\chi_{11}\psi_1+\chi_{12}\psi_2&=&0,\nonumber\\
\chi_{20}\psi_0+\chi_{21}\psi_1+\chi_{22}\psi_2+\chi_{23}\psi_3&=&0\nonumber\\\
...
\end{eqnarray}
where
\begin{eqnarray}
\chi_{00}&=&-\left(\frac{k^2}{r_+^{2p}}+m^2+\frac{i\omega}{r_+}p(d-1)\right)\nonumber\\
\chi_{01}&=&r_+^{2p}f^{\prime}(r_+)-2i\omega\nonumber\\
\chi_{10}&=&0\nonumber\\
\chi_{11}&=&2r_+^{2p}f^{\prime\prime}(r_+)+p(d+1)r_+^{2p-1}f^{\prime}(r_+)\nonumber\\
&&-\left(\frac{k^2}{r_+^{2p}}+m^2+\frac{i\omega}{r_+}p(d-1)\right)\nonumber\\
\chi_{12}&=&4r_+^{2p}f^{\prime}(r_+)-4i\omega\nonumber\\
.\nonumber\\
.\nonumber\\
.	
\end{eqnarray}
The coefficients $\chi_{ij}$ can arranged to form a square a matrix
$$\mathcal{M}=
\begin{bmatrix}
\chi_{00} & \chi_{01} & 0 & 0 & ...\\
\chi_{10} & \chi_{11} & \chi_{12} & 0& ...\\
\chi_{20} & \chi_{21} & \chi_{22} & \chi_{23}&...\\
...&...&...&...\\
\end{bmatrix}
$$
The pole-skipping points are to be obtained by simultaneously solving the equations \cite{Blake:2019otz}
\begin{eqnarray}
\chi_{n-1~n}=0;~\det(\mathcal{M})=0~.
\end{eqnarray}
\begin{widetext}
	\noindent We now provide the locations of some of these pole-skipping points. These read 
	\begin{eqnarray}
	\omega_1 = -i2\pi T_H;&&~k_1=-i\frac{2\pi T_H}{\sqrt{\frac{c}{2p(d-1)}}}\left[1+\frac{m^2}{2p(d-1)\pi}\left(\frac{4\pi}{c}\right)^{\frac{1}{(2p-1)}}T_H^{\frac{2-2p}{2p-1}}\right]^{1/2},\nonumber\\
	\omega_2 = -i4\pi T_H;&&~k_2=-i\frac{4\pi T_H}{\sqrt{\frac{c}{p(d-1)+2(c+1)-p(d+1)}}}\nonumber\\
	&&\times\left[1+\frac{m^2}{\left(4\pi\right)^2}\left(\frac{c}{p(d-1)+2(c+1)-p(d+1)}\right)\left(\frac{4\pi}{c}\right)^{\frac{2p}{2p-1}}T_H^{\frac{2-2p}{2p-1}}\right]^{1/2},\nonumber\\
	.\nonumber\\
	.\nonumber\\
	.\nonumber
	\end{eqnarray}
\end{widetext}
From the above one can observe that the pole-skipping points reside in the lower half of the complex $(\omega,k)$ plane. This is quite expected from the relation given in eq.\eqref{eq7} as we have considered scalar field field fluctuation which has spin $s=0$. These points are not related to the parameters of chaos as they are negative.		
\section{Acknowledgements}
\noindent AS would like to thank S.N. Bose National Centre for Basic Sciences for the financial support through its Advanced Postdoctoral Research Programme. The authors would like to thank the referees for valuable comments and suggestions.
\bibliographystyle{hephys}   
\bibliography{Reference}

\begin{thebibliography}{10}
\newcommand{\enquote}[1]{``#1''}

\bibitem{Stockmann_1999}
H.-J. Stöckmann, {Quantum Chaos: An Introduction}, Cambridge University Press
  1999.

\bibitem{Ullmo2014}
D.~Ullmo and S.~Tomsovic, \enquote{{Introduction to quantum chaos, 2014}},
  \emph{http://www.lptms.u-psud.fr/membres/ullmo/Articles/eolss-ullmo-tomsovic.pdf}
  .

\bibitem{Larkin1969QuasiclassicalMI}
A.~I. Larkin and Y.~N. Ovchinnikov, \enquote{Quasiclassical Method in the
  Theory of Superconductivity},
  \href{https://api.semanticscholar.org/CorpusID:117608877}{\emph{Journal of
  Experimental and Theoretical Physics} }.

\bibitem{Shenker:2013pqa}
S.~H. Shenker and D.~Stanford, \enquote{{Black holes and the butterfly
  effect}}, \href{http://dx.doi.org/10.1007/JHEP03(2014)067}{\emph{JHEP}
  \textbf{03} (2014) 067}, \href{http://arxiv.org/abs/1306.0622}{{\tt
  arXiv:1306.0622 [hep-th]}}.

\bibitem{Roberts:2016wdl}
D.~A. Roberts and B.~Swingle, \enquote{{Lieb-Robinson Bound and the Butterfly
  Effect in Quantum Field Theories}},
  \href{http://dx.doi.org/10.1103/PhysRevLett.117.091602}{\emph{Phys. Rev.
  Lett.} \textbf{117[9]} (2016) 091602},
  \href{http://arxiv.org/abs/1603.09298}{{\tt arXiv:1603.09298 [hep-th]}}.

\bibitem{Lieb:1972wy}
E.~H. Lieb and D.~W. Robinson, \enquote{{The finite group velocity of quantum
  spin systems}}, \href{http://dx.doi.org/10.1007/BF01645779}{\emph{Commun.
  Math. Phys.} \textbf{28} (1972) 251}.

\bibitem{Shenker:2014cwa}
S.~H. Shenker and D.~Stanford, \enquote{{Stringy effects in scrambling}},
  \href{http://dx.doi.org/10.1007/JHEP05(2015)132}{\emph{JHEP} \textbf{05}
  (2015) 132}, \href{http://arxiv.org/abs/1412.6087}{{\tt arXiv:1412.6087
  [hep-th]}}.

\bibitem{Roberts:2014isa}
D.~A. Roberts, D.~Stanford and L.~Susskind, \enquote{{Localized shocks}},
  \href{http://dx.doi.org/10.1007/JHEP03(2015)051}{\emph{JHEP} \textbf{03}
  (2015) 051}, \href{http://arxiv.org/abs/1409.8180}{{\tt arXiv:1409.8180
  [hep-th]}}.

\bibitem{Stanford:2015owe}
D.~Stanford, \enquote{{Many-body chaos at weak coupling}},
  \href{http://dx.doi.org/10.1007/JHEP10(2016)009}{\emph{JHEP} \textbf{10}
  (2016) 009}, \href{http://arxiv.org/abs/1512.07687}{{\tt arXiv:1512.07687
  [hep-th]}}.

\bibitem{Maldacena:2015waa}
J.~Maldacena, S.~H. Shenker and D.~Stanford, \enquote{{A bound on chaos}},
  \href{http://dx.doi.org/10.1007/JHEP08(2016)106}{\emph{JHEP} \textbf{08}
  (2016) 106}, \href{http://arxiv.org/abs/1503.01409}{{\tt arXiv:1503.01409
  [hep-th]}}.

\bibitem{Jahnke:2018off}
V.~Jahnke, \enquote{{Recent developments in the holographic description of
  quantum chaos}}, \href{http://dx.doi.org/10.1155/2019/9632708}{\emph{Adv.
  High Energy Phys.} \textbf{2019} (2019) 9632708},
  \href{http://arxiv.org/abs/1811.06949}{{\tt arXiv:1811.06949 [hep-th]}}.

\bibitem{Sekino:2008he}
Y.~Sekino and L.~Susskind, \enquote{{Fast Scramblers}},
  \href{http://dx.doi.org/10.1088/1126-6708/2008/10/065}{\emph{JHEP}
  \textbf{10} (2008) 065}, \href{http://arxiv.org/abs/0808.2096}{{\tt
  arXiv:0808.2096 [hep-th]}}.

\bibitem{Lashkari:2011yi}
N.~Lashkari, D.~Stanford, M.~Hastings, T.~Osborne and P.~Hayden,
  \enquote{{Towards the Fast Scrambling Conjecture}},
  \href{http://dx.doi.org/10.1007/JHEP04(2013)022}{\emph{JHEP} \textbf{04}
  (2013) 022}, \href{http://arxiv.org/abs/1111.6580}{{\tt arXiv:1111.6580
  [hep-th]}}.

\bibitem{Maldacena:1997re}
J.~M. Maldacena, \enquote{{The Large N limit of superconformal field theories
  and supergravity}},
  \href{http://dx.doi.org/10.4310/ATMP.1998.v2.n2.a1}{\emph{Adv. Theor. Math.
  Phys.} \textbf{2} (1998) 231},
  \href{http://arxiv.org/abs/hep-th/9711200}{{\tt arXiv:hep-th/9711200}}.

\bibitem{Gubser:1998bc}
S.~S. Gubser, I.~R. Klebanov and A.~M. Polyakov, \enquote{{Gauge theory
  correlators from noncritical string theory}},
  \href{http://dx.doi.org/10.1016/S0370-2693(98)00377-3}{\emph{Phys. Lett. B}
  \textbf{428} (1998) 105}, \href{http://arxiv.org/abs/hep-th/9802109}{{\tt
  arXiv:hep-th/9802109}}.

\bibitem{Witten:1998qj}
E.~Witten, \enquote{{Anti-de Sitter space and holography}},
  \href{http://dx.doi.org/10.4310/ATMP.1998.v2.n2.a2}{\emph{Adv. Theor. Math.
  Phys.} \textbf{2} (1998) 253},
  \href{http://arxiv.org/abs/hep-th/9802150}{{\tt arXiv:hep-th/9802150}}.

\bibitem{Grozdanov:2018kkt}
S.~Grozdanov, \enquote{{On the connection between hydrodynamics and quantum
  chaos in holographic theories with stringy corrections}},
  \href{http://dx.doi.org/10.1007/JHEP01(2019)048}{\emph{JHEP} \textbf{01}
  (2019) 048}, \href{http://arxiv.org/abs/1811.09641}{{\tt arXiv:1811.09641
  [hep-th]}}.

\bibitem{Natsuume:2019vcv}
M.~Natsuume and T.~Okamura, \enquote{{Pole-skipping with finite-coupling
  corrections}},
  \href{http://dx.doi.org/10.1103/PhysRevD.100.126012}{\emph{Phys. Rev. D}
  \textbf{100[12]} (2019) 126012}, \href{http://arxiv.org/abs/1909.09168}{{\tt
  arXiv:1909.09168 [hep-th]}}.

\bibitem{Israel:1976ur}
W.~Israel, \enquote{{Thermo field dynamics of black holes}},
  \href{http://dx.doi.org/10.1016/0375-9601(76)90178-X}{\emph{Phys. Lett. A}
  \textbf{57} (1976) 107}.

\bibitem{Maldacena:2001kr}
J.~M. Maldacena, \enquote{{Eternal black holes in anti-de Sitter}},
  \href{http://dx.doi.org/10.1088/1126-6708/2003/04/021}{\emph{JHEP}
  \textbf{04} (2003) 021}, \href{http://arxiv.org/abs/hep-th/0106112}{{\tt
  arXiv:hep-th/0106112}}.

\bibitem{Morrison:2012iz}
I.~A. Morrison and M.~M. Roberts, \enquote{{Mutual information between
  thermo-field doubles and disconnected holographic boundaries}},
  \href{http://dx.doi.org/10.1007/JHEP07(2013)081}{\emph{JHEP} \textbf{07}
  (2013) 081}, \href{http://arxiv.org/abs/1211.2887}{{\tt arXiv:1211.2887
  [hep-th]}}.

\bibitem{Wolf:2007tdq}
M.~M. Wolf, F.~Verstraete, M.~B. Hastings and J.~I. Cirac, \enquote{{Area Laws
  in Quantum Systems: Mutual Information and Correlations}},
  \href{http://dx.doi.org/10.1103/PhysRevLett.100.070502}{\emph{Phys. Rev.
  Lett.} \textbf{100[7]} (2008) 070502},
  \href{http://arxiv.org/abs/0704.3906}{{\tt arXiv:0704.3906 [quant-ph]}}.

\bibitem{Hayden:2011ag}
P.~Hayden, M.~Headrick and A.~Maloney, \enquote{{Holographic Mutual Information
  is Monogamous}},
  \href{http://dx.doi.org/10.1103/PhysRevD.87.046003}{\emph{Phys. Rev. D}
  \textbf{87[4]} (2013) 046003}, \href{http://arxiv.org/abs/1107.2940}{{\tt
  arXiv:1107.2940 [hep-th]}}.

\bibitem{Fischler:2012uv}
W.~Fischler, A.~Kundu and S.~Kundu, \enquote{{Holographic Mutual Information at
  Finite Temperature}},
  \href{http://dx.doi.org/10.1103/PhysRevD.87.126012}{\emph{Phys. Rev. D}
  \textbf{87[12]} (2013) 126012}, \href{http://arxiv.org/abs/1212.4764}{{\tt
  arXiv:1212.4764 [hep-th]}}.

\bibitem{Dray:1984ha}
T.~Dray and G.~'t~Hooft, \enquote{{The Gravitational Shock Wave of a Massless
  Particle}}, \href{http://dx.doi.org/10.1016/0550-3213(85)90525-5}{\emph{Nucl.
  Phys. B} \textbf{253} (1985) 173}.

\bibitem{Sfetsos:1994xa}
K.~Sfetsos, \enquote{{On gravitational shock waves in curved space-times}},
  \href{http://dx.doi.org/10.1016/0550-3213(94)00573-W}{\emph{Nucl. Phys. B}
  \textbf{436} (1995) 721}, \href{http://arxiv.org/abs/hep-th/9408169}{{\tt
  arXiv:hep-th/9408169}}.

\bibitem{Leichenauer:2014nxa}
S.~Leichenauer, \enquote{{Disrupting Entanglement of Black Holes}},
  \href{http://dx.doi.org/10.1103/PhysRevD.90.046009}{\emph{Phys. Rev. D}
  \textbf{90[4]} (2014) 046009}, \href{http://arxiv.org/abs/1405.7365}{{\tt
  arXiv:1405.7365 [hep-th]}}.

\bibitem{Sircar:2016old}
N.~Sircar, J.~Sonnenschein and W.~Tangarife, \enquote{{Extending the scope of
  holographic mutual information and chaotic behavior}},
  \href{http://dx.doi.org/10.1007/JHEP05(2016)091}{\emph{JHEP} \textbf{05}
  (2016) 091}, \href{http://arxiv.org/abs/1602.07307}{{\tt arXiv:1602.07307
  [hep-th]}}.

\bibitem{Jahnke:2017iwi}
V.~Jahnke, \enquote{{Delocalizing entanglement of anisotropic black branes}},
  \href{http://dx.doi.org/10.1007/JHEP01(2018)102}{\emph{JHEP} \textbf{01}
  (2018) 102}, \href{http://arxiv.org/abs/1708.07243}{{\tt arXiv:1708.07243
  [hep-th]}}.

\bibitem{Huang:2018snb}
W.-H. Huang, \enquote{{Butterfly Velocity in Quadratic Gravity}},
  \href{http://dx.doi.org/10.1088/1361-6382/aadb32}{\emph{Class. Quant. Grav.}
  \textbf{35[19]} (2018) 195004}, \href{http://arxiv.org/abs/1804.05527}{{\tt
  arXiv:1804.05527 [hep-th]}}.

\bibitem{Avila:2018sqf}
D.~\'Avila, V.~Jahnke and L.~Pati\~no, \enquote{{Chaos, Diffusivity, and
  Spreading of Entanglement in Magnetic Branes, and the Strengthening of the
  Internal Interaction}},
  \href{http://dx.doi.org/10.1007/JHEP09(2018)131}{\emph{JHEP} \textbf{09}
  (2018) 131}, \href{http://arxiv.org/abs/1805.05351}{{\tt arXiv:1805.05351
  [hep-th]}}.

\bibitem{Fischler:2018kwt}
W.~Fischler, V.~Jahnke and J.~F. Pedraza, \enquote{{Chaos and entanglement
  spreading in a non-commutative gauge theory}},
  \href{http://dx.doi.org/10.1007/JHEP11(2018)072}{\emph{JHEP} \textbf{11}
  (2018) 072}, [Erratum: JHEP 02, 149 (2021)],
  \href{http://arxiv.org/abs/1808.10050}{{\tt arXiv:1808.10050 [hep-th]}}.

\bibitem{Ageev:2021xgy}
D.~S. Ageev, \enquote{{Butterflies dragging the jets: on the chaotic nature of
  holographic QCD}}, \href{http://arxiv.org/abs/2105.04589}{{\tt
  arXiv:2105.04589 [hep-th]}}.

\bibitem{Mahish:2022xjz}
S.~Mahish and K.~Sil, \enquote{{Quantum information scrambling and quantum
  chaos in little string theory}},
  \href{http://dx.doi.org/10.1007/JHEP08(2022)041}{\emph{JHEP} \textbf{08}
  (2022) 041}, \href{http://arxiv.org/abs/2202.05865}{{\tt arXiv:2202.05865
  [hep-th]}}.

\bibitem{Chakrabortty:2022kvq}
S.~Chakrabortty, H.~Hoshino, S.~Pant and K.~Sil, \enquote{{A holographic study
  of the characteristics of chaos and correlation in the presence of
  backreaction}},
  \href{http://dx.doi.org/10.1016/j.physletb.2023.137749}{\emph{Phys. Lett. B}
  \textbf{838} (2023) 137749}, \href{http://arxiv.org/abs/2206.12555}{{\tt
  arXiv:2206.12555 [hep-th]}}.

\bibitem{Hartman:2013qma}
T.~Hartman and J.~Maldacena, \enquote{{Time Evolution of Entanglement Entropy
  from Black Hole Interiors}},
  \href{http://dx.doi.org/10.1007/JHEP05(2013)014}{\emph{JHEP} \textbf{05}
  (2013) 014}, \href{http://arxiv.org/abs/1303.1080}{{\tt arXiv:1303.1080
  [hep-th]}}.

\bibitem{Hartman:2015apr}
T.~Hartman and N.~Afkhami-Jeddi, \enquote{{Speed Limits for Entanglement}},
  \href{http://arxiv.org/abs/1512.02695}{{\tt arXiv:1512.02695 [hep-th]}}.

\bibitem{Mezei:2016zxg}
M.~Mezei, \enquote{{On entanglement spreading from holography}},
  \href{http://dx.doi.org/10.1007/JHEP05(2017)064}{\emph{JHEP} \textbf{05}
  (2017) 064}, \href{http://arxiv.org/abs/1612.00082}{{\tt arXiv:1612.00082
  [hep-th]}}.

\bibitem{Calabrese:2005in}
P.~Calabrese and J.~L. Cardy, \enquote{{Evolution of entanglement entropy in
  one-dimensional systems}},
  \href{http://dx.doi.org/10.1088/1742-5468/2005/04/P04010}{\emph{J. Stat.
  Mech.} \textbf{0504} (2005) P04010},
  \href{http://arxiv.org/abs/cond-mat/0503393}{{\tt arXiv:cond-mat/0503393}}.

\bibitem{Cotler:2016acd}
J.~S. Cotler, M.~P. Hertzberg, M.~Mezei and M.~T. Mueller,
  \enquote{{Entanglement Growth after a Global Quench in Free Scalar Field
  Theory}}, \href{http://dx.doi.org/10.1007/JHEP11(2016)166}{\emph{JHEP}
  \textbf{11} (2016) 166}, \href{http://arxiv.org/abs/1609.00872}{{\tt
  arXiv:1609.00872 [hep-th]}}.

\bibitem{Liu:2013iza}
H.~Liu and S.~J. Suh, \enquote{{Entanglement Tsunami: Universal Scaling in
  Holographic Thermalization}},
  \href{http://dx.doi.org/10.1103/PhysRevLett.112.011601}{\emph{Phys. Rev.
  Lett.} \textbf{112} (2014) 011601},
  \href{http://arxiv.org/abs/1305.7244}{{\tt arXiv:1305.7244 [hep-th]}}.

\bibitem{Liu:2013qca}
H.~Liu and S.~J. Suh, \enquote{{Entanglement growth during thermalization in
  holographic systems}},
  \href{http://dx.doi.org/10.1103/PhysRevD.89.066012}{\emph{Phys. Rev. D}
  \textbf{89[6]} (2014) 066012}, \href{http://arxiv.org/abs/1311.1200}{{\tt
  arXiv:1311.1200 [hep-th]}}.

\bibitem{Mezei:2016wfz}
M.~Mezei and D.~Stanford, \enquote{{On entanglement spreading in chaotic
  systems}}, \href{http://dx.doi.org/10.1007/JHEP05(2017)065}{\emph{JHEP}
  \textbf{05} (2017) 065}, \href{http://arxiv.org/abs/1608.05101}{{\tt
  arXiv:1608.05101 [hep-th]}}.

\bibitem{Qi:2017ttv}
X.-L. Qi and Z.~Yang, \enquote{{Butterfly velocity and bulk causal structure}},
  \href{http://arxiv.org/abs/1705.01728}{{\tt arXiv:1705.01728 [hep-th]}}.

\bibitem{Blake:2018leo}
M.~Blake, R.~A. Davison, S.~Grozdanov and H.~Liu, \enquote{{Many-body chaos and
  energy dynamics in holography}},
  \href{http://dx.doi.org/10.1007/JHEP10(2018)035}{\emph{JHEP} \textbf{10}
  (2018) 035}, \href{http://arxiv.org/abs/1809.01169}{{\tt arXiv:1809.01169
  [hep-th]}}.

\bibitem{Blake:2019otz}
M.~Blake, R.~A. Davison and D.~Vegh, \enquote{{Horizon constraints on
  holographic Green\textquoteright{}s functions}},
  \href{http://dx.doi.org/10.1007/JHEP01(2020)077}{\emph{JHEP} \textbf{01}
  (2020) 077}, \href{http://arxiv.org/abs/1904.12883}{{\tt arXiv:1904.12883
  [hep-th]}}.

\bibitem{Natusuume}
M.~Natsuume, \href{http://dx.doi.org/10.1007/978-4-431-55441-7}{AdS/CFT Duality
  User Guide}, Springer Tokyo 2015.

\bibitem{nastase}
H.~Năstase, \href{http://dx.doi.org/10.1017/CBO9781316090954}{Introduction to
  the AdS/CFT Correspondence}, Cambridge University Press 2015.

\bibitem{Grozdanov:2017ajz}
S.~Grozdanov, K.~Schalm and V.~Scopelliti, \enquote{{Black hole scrambling from
  hydrodynamics}},
  \href{http://dx.doi.org/10.1103/PhysRevLett.120.231601}{\emph{Phys. Rev.
  Lett.} \textbf{120[23]} (2018) 231601},
  \href{http://arxiv.org/abs/1710.00921}{{\tt arXiv:1710.00921 [hep-th]}}.

\bibitem{Grozdanov:2019uhi}
S.~Grozdanov, P.~K. Kovtun, A.~O. Starinets and P.~Tadi\'c, \enquote{{The
  complex life of hydrodynamic modes}},
  \href{http://dx.doi.org/10.1007/JHEP11(2019)097}{\emph{JHEP} \textbf{11}
  (2019) 097}, \href{http://arxiv.org/abs/1904.12862}{{\tt arXiv:1904.12862
  [hep-th]}}.

\bibitem{Natsuume:2019sfp}
M.~Natsuume and T.~Okamura, \enquote{{Holographic chaos, pole-skipping, and
  regularity}}, \href{http://dx.doi.org/10.1093/ptep/ptz155}{\emph{PTEP}
  \textbf{2020[1]} (2020) 013B07}, \href{http://arxiv.org/abs/1905.12014}{{\tt
  arXiv:1905.12014 [hep-th]}}.

\bibitem{Natsuume:2019xcy}
M.~Natsuume and T.~Okamura, \enquote{{Nonuniqueness of Green\textquoteright{}s
  functions at special points}},
  \href{http://dx.doi.org/10.1007/JHEP12(2019)139}{\emph{JHEP} \textbf{12}
  (2019) 139}, \href{http://arxiv.org/abs/1905.12015}{{\tt arXiv:1905.12015
  [hep-th]}}.

\bibitem{Ceplak:2019ymw}
N.~Ceplak, K.~Ramdial and D.~Vegh, \enquote{{Fermionic pole-skipping in
  holography}}, \href{http://dx.doi.org/10.1007/JHEP07(2020)203}{\emph{JHEP}
  \textbf{07} (2020) 203}, \href{http://arxiv.org/abs/1910.02975}{{\tt
  arXiv:1910.02975 [hep-th]}}.

\bibitem{Ceplak:2021efc}
N.~Ceplak and D.~Vegh, \enquote{{Pole-skipping and Rarita-Schwinger fields}},
  \href{http://dx.doi.org/10.1103/PhysRevD.103.106009}{\emph{Phys. Rev. D}
  \textbf{103[10]} (2021) 106009}, \href{http://arxiv.org/abs/2101.01490}{{\tt
  arXiv:2101.01490 [hep-th]}}.

\bibitem{Wang:2022mcq}
D.~Wang and Z.-Y. Wang, \enquote{{Pole Skipping in Holographic Theories with
  Bosonic Fields}},
  \href{http://dx.doi.org/10.1103/PhysRevLett.129.231603}{\emph{Phys. Rev.
  Lett.} \textbf{129[23]} (2022) 231603},
  \href{http://arxiv.org/abs/2208.01047}{{\tt arXiv:2208.01047 [hep-th]}}.

\bibitem{Ning:2023ggs}
S.~Ning, D.~Wang and Z.-Y. Wang, \enquote{{Pole skipping in holographic
  theories with gauge and fermionic fields}},
  \href{http://dx.doi.org/10.1007/JHEP12(2023)084}{\emph{JHEP} \textbf{12}
  (2023) 084}, \href{http://arxiv.org/abs/2308.08191}{{\tt arXiv:2308.08191
  [hep-th]}}.

\bibitem{Blake:2017ris}
M.~Blake, H.~Lee and H.~Liu, \enquote{{A quantum hydrodynamical description for
  scrambling and many-body chaos}},
  \href{http://dx.doi.org/10.1007/JHEP10(2018)127}{\emph{JHEP} \textbf{10}
  (2018) 127}, \href{http://arxiv.org/abs/1801.00010}{{\tt arXiv:1801.00010
  [hep-th]}}.

\bibitem{Choi:2020tdj}
C.~Choi, M.~Mezei and G.~S\'arosi, \enquote{{Pole skipping away from maximal
  chaos}}, \href{http://arxiv.org/abs/2010.08558}{{\tt arXiv:2010.08558
  [hep-th]}}.

\bibitem{Abbasi:2019rhy}
N.~Abbasi and J.~Tabatabaei, \enquote{{Quantum chaos, pole-skipping and
  hydrodynamics in a holographic system with chiral anomaly}},
  \href{http://dx.doi.org/10.1007/JHEP03(2020)050}{\emph{JHEP} \textbf{03}
  (2020) 050}, \href{http://arxiv.org/abs/1910.13696}{{\tt arXiv:1910.13696
  [hep-th]}}.

\bibitem{Ahn:2020bks}
Y.~Ahn, V.~Jahnke, H.-S. Jeong, K.-Y. Kim, K.-S. Lee and M.~Nishida,
  \enquote{{Pole-skipping of scalar and vector fields in hyperbolic space:
  conformal blocks and holography}},
  \href{http://dx.doi.org/10.1007/JHEP09(2020)111}{\emph{JHEP} \textbf{09}
  (2020) 111}, \href{http://arxiv.org/abs/2006.00974}{{\tt arXiv:2006.00974
  [hep-th]}}.

\bibitem{Ahn:2020baf}
Y.~Ahn, V.~Jahnke, H.-S. Jeong, K.-Y. Kim, K.-S. Lee and M.~Nishida,
  \enquote{{Classifying pole-skipping points}},
  \href{http://dx.doi.org/10.1007/JHEP03(2021)175}{\emph{JHEP} \textbf{03}
  (2021) 175}, \href{http://arxiv.org/abs/2010.16166}{{\tt arXiv:2010.16166
  [hep-th]}}.

\bibitem{Kim:2020url}
K.-Y. Kim, K.-S. Lee and M.~Nishida, \enquote{{Holographic scalar and vector
  exchange in OTOCs and pole-skipping phenomena}},
  \href{http://dx.doi.org/10.1007/JHEP04(2021)092}{\emph{JHEP} \textbf{04}
  (2021) 092}, [Erratum: JHEP 04, 229 (2021)],
  \href{http://arxiv.org/abs/2011.13716}{{\tt arXiv:2011.13716 [hep-th]}}.

\bibitem{Natsuume:2020snz}
M.~Natsuume and T.~Okamura, \enquote{{Pole-skipping and zero temperature}},
  \href{http://dx.doi.org/10.1103/PhysRevD.103.066017}{\emph{Phys. Rev. D}
  \textbf{103[6]} (2021) 066017}, \href{http://arxiv.org/abs/2011.10093}{{\tt
  arXiv:2011.10093 [hep-th]}}.

\bibitem{Abbasi:2020xli}
N.~Abbasi and M.~Kaminski, \enquote{{Constraints on quasinormal modes and
  bounds for critical points from pole-skipping}},
  \href{http://dx.doi.org/10.1007/JHEP03(2021)265}{\emph{JHEP} \textbf{03}
  (2021) 265}, \href{http://arxiv.org/abs/2012.15820}{{\tt arXiv:2012.15820
  [hep-th]}}.

\bibitem{Ramirez:2020qer}
D.~M. Ramirez, \enquote{{Chaos and pole skipping in CFT$_{2}$}},
  \href{http://dx.doi.org/10.1007/JHEP12(2021)006}{\emph{JHEP} \textbf{12}
  (2021) 006}, \href{http://arxiv.org/abs/2009.00500}{{\tt arXiv:2009.00500
  [hep-th]}}.

\bibitem{Abbasi:2020ykq}
N.~Abbasi and S.~Tahery, \enquote{{Complexified quasinormal modes and the
  pole-skipping in a holographic system at finite chemical potential}},
  \href{http://dx.doi.org/10.1007/JHEP10(2020)076}{\emph{JHEP} \textbf{10}
  (2020) 076}, \href{http://arxiv.org/abs/2007.10024}{{\tt arXiv:2007.10024
  [hep-th]}}.

\bibitem{Sil:2020jhr}
K.~Sil, \enquote{{Pole skipping and chaos in anisotropic plasma: a holographic
  study}}, \href{http://dx.doi.org/10.1007/JHEP03(2021)232}{\emph{JHEP}
  \textbf{03} (2021) 232}, \href{http://arxiv.org/abs/2012.07710}{{\tt
  arXiv:2012.07710 [hep-th]}}.

\bibitem{Yuan:2020fvv}
H.~Yuan and X.-H. Ge, \enquote{{Pole-skipping and hydrodynamic analysis in
  Lifshitz, AdS$_{2}$ and Rindler geometries}},
  \href{http://dx.doi.org/10.1007/JHEP06(2021)165}{\emph{JHEP} \textbf{06}
  (2021) 165}, \href{http://arxiv.org/abs/2012.15396}{{\tt arXiv:2012.15396
  [hep-th]}}.

\bibitem{Blake:2021hjj}
M.~Blake and R.~A. Davison, \enquote{{Chaos and pole-skipping in rotating black
  holes}}, \href{http://dx.doi.org/10.1007/JHEP01(2022)013}{\emph{JHEP}
  \textbf{01} (2022) 013}, \href{http://arxiv.org/abs/2111.11093}{{\tt
  arXiv:2111.11093 [hep-th]}}.

\bibitem{Yuan:2021ets}
H.~Yuan and X.-H. Ge, \enquote{{Analogue of the pole-skipping phenomenon in
  acoustic black holes}},
  \href{http://dx.doi.org/10.1140/epjc/s10052-022-10129-y}{\emph{Eur. Phys. J.
  C} \textbf{82[2]} (2022) 167}, \href{http://arxiv.org/abs/2110.08074}{{\tt
  arXiv:2110.08074 [hep-th]}}.

\bibitem{Yuan:2023tft}
H.~Yuan, X.-H. Ge, K.-Y. Kim, C.-W. Ji and Y.~Ahn, \enquote{{Pole-skipping
  points in 2D gravity and SYK model}},
  \href{http://dx.doi.org/10.1007/JHEP08(2023)157}{\emph{JHEP} \textbf{08}
  (2023) 157}, \href{http://arxiv.org/abs/2303.04801}{{\tt arXiv:2303.04801
  [hep-th]}}.

\bibitem{Yadav:2023hyg}
G.~Yadav, S.~S. Kushwah and A.~Misra, \enquote{{Pole-Skipping and Chaos in Too
  ${\cal M}$UCH Hot QCD}}, \href{http://arxiv.org/abs/2311.09306}{{\tt
  arXiv:2311.09306 [hep-th]}}.

\bibitem{Ahn:2019rnq}
Y.~Ahn, V.~Jahnke, H.-S. Jeong and K.-Y. Kim, \enquote{{Scrambling in
  Hyperbolic Black Holes: shock waves and pole-skipping}},
  \href{http://dx.doi.org/10.1007/JHEP10(2019)257}{\emph{JHEP} \textbf{10}
  (2019) 257}, \href{http://arxiv.org/abs/1907.08030}{{\tt arXiv:1907.08030
  [hep-th]}}.

\bibitem{Jeong:2021zhz}
H.-S. Jeong, K.-Y. Kim and Y.-W. Sun, \enquote{{Bound of diffusion constants
  from pole-skipping points: spontaneous symmetry breaking and magnetic
  field}}, \href{http://dx.doi.org/10.1007/JHEP07(2021)105}{\emph{JHEP}
  \textbf{07} (2021) 105}, \href{http://arxiv.org/abs/2104.13084}{{\tt
  arXiv:2104.13084 [hep-th]}}.

\bibitem{Jeong:2022luo}
H.-S. Jeong, K.-Y. Kim and Y.-W. Sun, \enquote{{Quasi-normal modes of dyonic
  black holes and magneto-hydrodynamics}},
  \href{http://dx.doi.org/10.1007/JHEP07(2022)065}{\emph{JHEP} \textbf{07}
  (2022) 065}, \href{http://arxiv.org/abs/2203.02642}{{\tt arXiv:2203.02642
  [hep-th]}}.

\bibitem{Jeong:2023rck}
H.-S. Jeong, C.-W. Ji and K.-Y. Kim, \enquote{{Pole-skipping in rotating BTZ
  black holes}}, \href{http://dx.doi.org/10.1007/JHEP08(2023)139}{\emph{JHEP}
  \textbf{08} (2023) 139}, \href{http://arxiv.org/abs/2306.14805}{{\tt
  arXiv:2306.14805 [hep-th]}}.

\bibitem{Jeong:2023ynk}
H.-S. Jeong, \enquote{{Quantum chaos and pole-skipping in a semilocally
  critical IR fixed point}}, \href{http://arxiv.org/abs/2309.13412}{{\tt
  arXiv:2309.13412 [hep-th]}}.

\bibitem{Kulkarni:2012re}
S.~Kulkarni, B.-H. Lee, C.~Park and R.~Roychowdhury, \enquote{{Non-conformal
  Hydrodynamics in Einstein-dilaton Theory}},
  \href{http://dx.doi.org/10.1007/JHEP09(2012)004}{\emph{JHEP} \textbf{09}
  (2012) 004}, \href{http://arxiv.org/abs/1205.3883}{{\tt arXiv:1205.3883
  [hep-th]}}.

\bibitem{Kulkarni:2012in}
S.~Kulkarni, B.-H. Lee, J.-H. Oh, C.~Park and R.~Roychowdhury,
  \enquote{{Transports in non-conformal holographic fluids}},
  \href{http://dx.doi.org/10.1007/JHEP03(2013)149}{\emph{JHEP} \textbf{03}
  (2013) 149}, \href{http://arxiv.org/abs/1211.5972}{{\tt arXiv:1211.5972
  [hep-th]}}.

\bibitem{Park:2012lzs}
C.~Park, \enquote{{Holographic Aspects of a Relativistic Nonconformal Theory}},
  \href{http://dx.doi.org/10.1155/2013/389541}{\emph{Adv. High Energy Phys.}
  \textbf{2013} (2013) 389541}, \href{http://arxiv.org/abs/1209.0842}{{\tt
  arXiv:1209.0842 [hep-th]}}.

\bibitem{Park:2015afa}
C.~Park, \enquote{{Holographic entanglement entropy in the nonconformal
  medium}}, \href{http://dx.doi.org/10.1103/PhysRevD.91.126003}{\emph{Phys.
  Rev. D} \textbf{91[12]} (2015) 126003},
  \href{http://arxiv.org/abs/1501.02908}{{\tt arXiv:1501.02908 [hep-th]}}.

\bibitem{Charmousis:2009xr}
C.~Charmousis, B.~Gouteraux and J.~Soda, \enquote{{Einstein-Maxwell-Dilaton
  theories with a Liouville potential}},
  \href{http://dx.doi.org/10.1103/PhysRevD.80.024028}{\emph{Phys. Rev. D}
  \textbf{80} (2009) 024028}, \href{http://arxiv.org/abs/0905.3337}{{\tt
  arXiv:0905.3337 [gr-qc]}}.

\bibitem{Gubser:2000nd}
S.~S. Gubser, \enquote{{Curvature singularities: The Good, the bad, and the
  naked}}, \href{http://dx.doi.org/10.4310/ATMP.2000.v4.n3.a6}{\emph{Adv.
  Theor. Math. Phys.} \textbf{4} (2000) 679},
  \href{http://arxiv.org/abs/hep-th/0002160}{{\tt arXiv:hep-th/0002160}}.

\bibitem{Gouteraux:2011ce}
B.~Gouteraux and E.~Kiritsis, \enquote{{Generalized Holographic Quantum
  Criticality at Finite Density}},
  \href{http://dx.doi.org/10.1007/JHEP12(2011)036}{\emph{JHEP} \textbf{12}
  (2011) 036}, \href{http://arxiv.org/abs/1107.2116}{{\tt arXiv:1107.2116
  [hep-th]}}.

\bibitem{Saha:2020fon}
A.~Saha, S.~Gangopadhyay and J.~P. Saha, \enquote{{Generalized entanglement
  temperature and entanglement Smarr relation}},
  \href{http://dx.doi.org/10.1103/PhysRevD.102.086010}{\emph{Phys. Rev. D}
  \textbf{102[8]} (2020) 086010}, \href{http://arxiv.org/abs/2004.00867}{{\tt
  arXiv:2004.00867 [hep-th]}}.

\bibitem{Chew:2022enh}
X.~Y. Chew, D.-h. Yeom and J.~L. Bl\'azquez-Salcedo, \enquote{{Properties of
  scalar hairy black holes and scalarons with asymmetric potential}},
  \href{http://dx.doi.org/10.1103/PhysRevD.108.044020}{\emph{Phys. Rev. D}
  \textbf{108[4]} (2023) 044020}, \href{http://arxiv.org/abs/2210.01313}{{\tt
  arXiv:2210.01313 [gr-qc]}}.

\bibitem{Chew:2023olq}
X.~Y. Chew and K.-G. Lim, \enquote{{Scalar Hairy Black Holes with Inverted
  Mexican Hat Potential}}, \href{http://arxiv.org/abs/2307.13972}{{\tt
  arXiv:2307.13972 [gr-qc]}}.

\bibitem{Shenker:2013yza}
S.~H. Shenker and D.~Stanford, \enquote{{Multiple Shocks}},
  \href{http://dx.doi.org/10.1007/JHEP12(2014)046}{\emph{JHEP} \textbf{12}
  (2014) 046}, \href{http://arxiv.org/abs/1312.3296}{{\tt arXiv:1312.3296
  [hep-th]}}.

\bibitem{Ryu:2006bv}
S.~Ryu and T.~Takayanagi, \enquote{{Holographic derivation of entanglement
  entropy from AdS/CFT}},
  \href{http://dx.doi.org/10.1103/PhysRevLett.96.181602}{\emph{Phys. Rev.
  Lett.} \textbf{96} (2006) 181602},
  \href{http://arxiv.org/abs/hep-th/0603001}{{\tt arXiv:hep-th/0603001}}.

\bibitem{Ryu:2006ef}
S.~Ryu and T.~Takayanagi, \enquote{{Aspects of Holographic Entanglement
  Entropy}}, \href{http://dx.doi.org/10.1088/1126-6708/2006/08/045}{\emph{JHEP}
  \textbf{08} (2006) 045}, \href{http://arxiv.org/abs/hep-th/0605073}{{\tt
  arXiv:hep-th/0605073}}.

\bibitem{Hubeny:2007xt}
V.~E. Hubeny, M.~Rangamani and T.~Takayanagi, \enquote{{A Covariant holographic
  entanglement entropy proposal}},
  \href{http://dx.doi.org/10.1088/1126-6708/2007/07/062}{\emph{JHEP}
  \textbf{07} (2007) 062}, \href{http://arxiv.org/abs/0705.0016}{{\tt
  arXiv:0705.0016 [hep-th]}}.

\bibitem{Kundu:2021nwp}
A.~Kundu, \enquote{{Wormholes and holography: an introduction}},
  \href{http://dx.doi.org/10.1140/epjc/s10052-022-10376-z}{\emph{Eur. Phys. J.
  C} \textbf{82[5]} (2022) 447}, \href{http://arxiv.org/abs/2110.14958}{{\tt
  arXiv:2110.14958 [hep-th]}}.

\bibitem{Hubeny:2012ry}
V.~E. Hubeny, \enquote{{Extremal surfaces as bulk probes in AdS/CFT}},
  \href{http://dx.doi.org/10.1007/JHEP07(2012)093}{\emph{JHEP} \textbf{07}
  (2012) 093}, \href{http://arxiv.org/abs/1203.1044}{{\tt arXiv:1203.1044
  [hep-th]}}.

\end{thebibliography}

\end{document}